\documentclass[conference]{IEEEtran}
\IEEEoverridecommandlockouts
\usepackage{graphicx}
\usepackage{subcaption}
\usepackage{float}
\usepackage{todonotes}
\usepackage{tabularx}
\usepackage{booktabs}
\usepackage{tcolorbox}
\usepackage{enumitem}
\usepackage{tikz}
\usepackage{adjustbox}
\usepackage{stfloats}
\usetikzlibrary{positioning,fit}
\newtcolorbox{observationbox}{colback=red!5!white,colframe=red!75!black}
\usepackage{cite}
\usepackage{amsmath,amssymb,amsfonts}
\usepackage{textcomp}
\usepackage[hyphens]{url}
\def\BibTeX{{\rm B\kern-.05em{\sc i\kern-.025em b}\kern-.08em
    T\kern-.1667em\lower.7ex\hbox{E}\kern-.125emX}}
\begin{document}
\pdfpagewidth=8.5in
\pdfpageheight=11in

\pagenumbering{arabic}

\title{Understanding Inference Scaling for LLMs: Bottlenecks, Trade-offs, and Performance Principles}
\author{
    \IEEEauthorblockN{
        Moiz Arif\IEEEauthorrefmark{1},
        Avinash Maurya\IEEEauthorrefmark{2},
        Sudharshan Vazhkudai\IEEEauthorrefmark{1},
        Bogdan Nicolae\IEEEauthorrefmark{2}
    }
    \IEEEauthorblockA{
        \IEEEauthorrefmark{1}\it{Micron Technology Inc., Austin, TX, USA} \\
        \IEEEauthorrefmark{2}\it{Argonne National Laboratory, Lemont, IL, USA} \\
        Email: {\IEEEauthorrefmark{1}\{marifa, svazhkudai\}@micron.com; \IEEEauthorrefmark{2}\{amaurya, bnicolae\}@anl.gov}
    }
}

\maketitle

\thispagestyle{plain}
\pagestyle{plain}

\begin{abstract}
The transition from standard generative AI to \emph{reasoning-centric architectures}, exemplified by models capable of extensive Chain-of-Thought~(CoT) processing, marks a fundamental paradigm shift in system requirements. Unlike traditional workloads dominated by compute-bound prefill, reasoning workloads generate long chains of reasoning tokens that shift inference into a \emph{Capacity-Bound regime}. This paper presents a comprehensive system characterization, evaluating models ranging from 8B to 671B parameters on GPUs clusters. By systematically exploring the interplay between Data, Tensor, and Pipeline parallelism, we identify critical bottlenecks that defy standard scaling heuristics. 
Our analysis reveals that 
data parallelism is throughput efficient for small models but hits a capacity trap on reasoning workloads as KV-cache fragmentation forces early throttling resulting in sub-optimal compute utilization. Tensor parallelism unlocks stranded memory and delivers sublinear gains near the 32B crossover. At frontier scale, dense models (e.g., Llama-405B) are interconnect and memory-bandwidth bound and favor high-degree TP, while sparse Mixture-of-Experts (MoE) models (e.g., DeepSeek-R1) are limited by routing and synchronization latency and benefit from hybrid strategies. These insights provide a rigorous decision framework for navigating the reasoning cliff, establishing new architectural imperatives for the next generation of inference infrastructure.
\end{abstract}


\section{Introduction}
\label{sec:intro}
\subsection{Motivation}
The deployment of Large Language Models~(LLMs) has transitioned from a capability demonstration to a cornerstone of industrial computing. However, the emergence of \emph{reasoning-centric architectures}, exemplified by DeepSeek-R1~\cite{deepseekai2025deepseekr1} and OpenAI's o1~\cite{openai2024o1} marks a fundamental paradigm shift in system requirements. Unlike standard generative models that prioritize fluent text generation, reasoning models employ CoT\cite{wei2022chainofthought} like processing to generate extensive intermediate logic traces before producing a final answer.

This shift fundamentally alters the resource signature of AI inference. While traditional LLM serving is often dominated by compute-bound prefill or bandwidth-bound decoding of short responses, reasoning workloads typically demonstrate a \emph{Capacity-Bound} inference regime. The generation of thousands of intermediate ``thinking'' tokens~\cite{wang2025attentionheads} creates massive, persistent Key-Value (KV) cache footprints that saturate High Bandwidth Memory (HBM) long before compute utilization peaks~\cite{KVCacheProfile-IPDPS25}. Consequently, standard scaling heuristics such as maximizing batch size for throughput or relying solely on Data Parallelism~(DP) become counter-productive, triggering severe memory thrashing, scheduler preemption, and non-linear latency spikes~\cite{agrawal2024vidur,li2024llm}.

Deploying these frontier-scale models (often exceeding 400B parameters~\cite{dubey2024llama3herdmodels,deepseekai2024deepseekv3}) requires navigating a complex, high-dimensional design space~\cite{isaev2023calculon}. System architects must balance the low-latency requirements of interactive reasoning against the strict capacity limits of modern accelerators. This necessitates a move beyond monolithic scaling toward nuanced combinations of Data, Tensor~\cite{shoeybi2019megatron}, and Pipeline parallelism~\cite{huang2019gpipe}, tailored specifically to the sparsity and sequence length characteristics of the model~\cite{zhao2025insights}.

\subsection{System Challenges in Reasoning Inference}
Deploying reasoning-centric models at scale introduces distinct system-level bottlenecks that are structurally different from training~\cite{10.1145/3545008.3545054} or standard chat inference:

\begin{itemize}[leftmargin=*]
    \item \textbf{The Capacity Wall:} Reasoning traces with Output Sequence Length~($OSL) \gg 10k$ cause the KV cache to grow linearly, rapidly exhausting per-GPU HBM~\cite{li2024survey}. In DP with engines such as vLLM~\cite{kwon2023vllm}, this leads to ``stranded capacity'' where fragmented memory across replicas forces premature request throttling despite low compute utilization~\cite{KVCacheProfile-IPDPS25}.
    \item \textbf{The Parallelism Efficiency Gap:} While Tensor Parallelism~(TP) alleviates memory pressure by using collective HBM of multiple GPUs, it incurs communication overheads that degrade performance beyond intra-node scaling~\cite{chitty2024llm}. Conversely, Pipeline Parallelism~(PP) offers memory relief but suffers from ``pipeline bubbles'' that are difficult to hide in serial, auto-regressive generation tasks~\cite{agrawal2023sarathi}.
    \item \textbf{Architectural Divergence:} The optimal strategy is no longer universal as we show in our analysis. As we demonstrate, dense models (e.g., Llama-3.1-405B) behave fundamentally differently from sparse MoE models (e.g., DeepSeek-R1-671B), requiring bespoke parallelization hierarchies to align with their respective compute-to-communication ratios~\cite{cao2025moe}.
\end{itemize}

While recent work has improved kernel efficiency and KV management, the dominant scaling limits for reasoning workloads increasingly arise from system-level capacity and scheduling dynamics rather than operator inefficiencies. This work isolates and quantifies these limits under realistic inference conditions.

\subsection{Contributions}
This paper presents a systematic characterization of GPU-based inference for reasoning-centric LLMs. We evaluate models ranging from 8B to 671B parameters on a cluster of NVIDIA H200 GPUs, identifying the breakpoints where standard scaling laws fall short.
Our specific contributions are:
\begin{enumerate}[leftmargin=*]
    \item \textbf{Characterization of the ``Parallelism Transition Point'':} We empirically determine the model size and sequence length thresholds where DP collapses due to KV saturation, necessitating a transition to hybrid or TP approaches.
    \item \textbf{Quantification of the ``Reasoning Gap'':} We demonstrate how the extended decoding phases of reasoning models shift the critical bottleneck from prefill compute (TTFT) to decode memory capacity (TPOT).
    \item \textbf{Evaluation of Frontier Architectures:} We provide a comparative analysis of large dense versus sparse scaling, revealing that MoE architectures favor hybrid strategies (PP+TP) to mitigate synchronization costs, whereas dense models demand high-bandwidth TP configurations.
    \item \textbf{Operational Guidance:} Based on these insights, we provide guidance for selecting optimal parallelism strategies that maximize fleet-level throughput while meeting strict latency SLAs for reasoning tasks.
\end{enumerate}

\section{Background and Related Work}
\label{sec:background}

\subsection{Reasoning Inference Breakdown}
Reasoning-centric inference is distinct from standard chat workloads in its execution flow and resource demands. It proceeds in two phases with orthogonal hardware requirements:

\paragraph*{\bf Prefill (The Compute Phase)}
The model processes the user prompt (input sequence) in parallel and dominated by large matrix-matrix multiplications (GEMMs) and is strictly compute-bound. Modern GPUs effectively utilize their tensor cores, achieving high Streaming Multiprocessor (SM) occupancy. The latency of this phase determines the TTFT.

\paragraph*{\bf Decode (The Memory Phase)}
Following prefill, the model enters the autoregressive generation loop. Unlike standard chat ($OSL \approx 500$), reasoning traces can exceed 10000 tokens~\cite{xu2025towards}. Each token generation requires reading the entire model weight set and the active KV cache from HBM~\cite{KVCacheProfile-IPDPS25}. This phase is strictly \emph{bandwidth-bound}; the arithmetic intensity collapses, and the GPU spends the majority of cycles performing memory reads. Our telemetry indicates that for reasoning workloads, the system spends $>99\%$ of its wall-clock time in this inefficient regime~\cite{agrawal2024taming}.

\subsection{Model Architectures and Attention Mechanisms}
The impact of the GPU memory exhaustion is modulated by the specific architecture of the model. We characterize two distinct paradigms found in state-of-the-art reasoning engines:

\paragraph*{\bf Dense Architectures (Grouped-Query Attention)}
Models like \emph{Llama-3.1-405B}~\cite{dubey2024llama3herdmodels} utilize a dense transformer architecture where every parameter is active for every token. To mitigate memory pressure, these models employ \emph{Grouped-Query Attention (GQA)} (typically 8 KV heads)~\cite{ainslie2023gqa}, which reduces the KV footprint by $3\times$--$8\times$ compared to standard Multi-Head Attention (MHA)~\cite{cordonnier2020multi}. However, the memory cost remains linear with layer count~\cite{alizadeh2024llm}. For the 405B model ($\approx$126 layers), the KV cache consumes $\approx$1.05 MB per token in FP16. Serving a batch of 128 requests with 10k reasoning tokens each consumes over 1.3 TB of memory solely for the cache, far outstripping the capacity of a single H200 GPU.

\paragraph*{\bf Sparse Architectures (Multi-Head Latent Attention)}
The \emph{DeepSeek-R1-671B}~\cite{deepseekai2025deepseekr1} model utilizes a MoE architecture, activating only $\approx$37B parameters per token. Crucially, it employs \emph{Multi-Head Latent Attention (MLA)}, which compresses the KV cache into a low-rank latent vector. This architectural choice effectively decouples the KV cache size from the number of attention heads, allowing R1 to sustain long reasoning contexts with a relatively lower memory footprint than dense models of same scale~\cite{liu2024minicache}.

\subsection{Memory Hierarchy and Scheduling}
To manage these massive footprints, we utilize \emph{vLLM}~\cite{kwon2023vllm}, an advanced, widely-used inference engine with \emph{PagedAttention}, which partitions the KV cache into non-contiguous blocks to eliminate internal fragmentation. The scheduler plays a critical role in this architecture:
\begin{itemize}[leftmargin=*]
    \item \textbf{Chunked Prefill:} To maximize GPU throughput, the scheduler splits long input sequences into smaller chunks~\cite{agrawal2023sarathi}. These chunks are processed iteratively and added greedily to batches, allowing the system to interleave prefill computation with ongoing decode requests. This smooths SM occupancy and mitigates head-of-line blocking, though it can fragment memory traffic.
   \item \textbf{Preemption:} When the ``Reasoning Cliff'' is reached (HBM saturation), the scheduler must preempt active requests~\cite{KVCacheProfile-IPDPS25}, moving them to a ``Waiting'' queue or swapping them to CPU host memory. This is a defensive mechanism to prevent OOM but triggers a re-computation or swapping penalty that degrades tail latency.
\end{itemize}

We focus on HBM-resident KV caches to isolate fundamental capacity and bandwidth limits of reasoning-centric inference without introducing latency trade-offs from disaggregated memory tiers. Techniques such as KV offloading~\cite{Offload-SC25}, prefetching~\cite{dong2026accelerating}, quantization~\cite{PagedEviction-EACL26}, and compression expand the optimization space but are ultimately constrained along the decode path by latency, bandwidth, and scheduling overheads. We therefore treat these techniques as complementary and orthogonal to our analysis, which characterizes when KV-cache pressure dominates even under HBM-only conditions.

\subsection{Parallelism Taxonomy}
Scaling these models requires distributing computation across several GPUs. We evaluate four strategies:

\paragraph*{\bf Data Parallelism (DP)}
The model is replicated on each GPU. While efficient for throughput, DP suffers from \emph{KV Fragmentation} as each replica holds a full copy of weights (e.g., $\approx$800GB for 405B which exceeds single GPU memory capacity and even for smaller models, weight replication reduces available KV space).

\paragraph*{\bf Tensor Parallelism (TP)}
Individual layers are sharded across GPUs. TP uses the aggregated HBM capacity without unnecessary model replication. However, it introduces high-frequency all-Reduce communication at every layer~\cite{agrawal2024vidur}. For the 32B model, we observe a transition where the benefit of aggregate memory outperforms the cost of communication.

\paragraph*{\bf Pipeline Parallelism (PP)}
Layers are partitioned sequentially across GPUs. PP reduces memory footprint without the high communication frequency of TP. However, it introduces ``Pipeline Bubbles'' (idle time)~\cite{huang2019gpipe} due to lack of sufficient concurrent requests to fill these bubbles without incurring unacceptable queueing latency.

\paragraph*{\bf Hybrid Parallelism}
A hierarchical combination (e.g., TP within a node, PP across nodes). Our analysis suggests this is critical for frontier sparse models to balance memory pooling with synchronization overheads.

\section{Experimental Methodology}
\label{sec:methodology}

\subsection{Testbed and Inference Engine}
All experiments are performed on a high-performance inference node equipped with 8$\times$ NVIDIA H200 Tensor Core GPUs (SXM5 form factor). Each H200 GPU features 141~GB of HBM3e memory providing a peak memory bandwidth of 4.8~TB/s, and delivers a theoretical peak of 1,979 TFLOPS for FP16/BF16 tensor operations. The GPUs are fully interconnected via fourth-generation NVLink and NVSwitch, providing a bidirectional GPU-to-GPU bandwidth of 900~GB/s per GPU, facilitating efficient all-reduce operations essential for TP. The host system is driven by dual Intel Xeon Platinum 8558P processors with 2TB DDR5 system memory.

We employ the vLLM v1 inference engine, leveraging its PagedAttention mechanism to mitigate memory fragmentation. The engine is configured with a block size of $B=16$ to balance quantization granularity with memory access efficiency. We utilize the default First-Come-First-Served (FCFS) scheduling policy but tune the \texttt{max\_num\_batched\_tokens} and \texttt{max\_num\_seqs} parameters to characterize the concurrency limits.

Our evaluation focuses on system-level behavior within a single NVLink-connected 8-GPU node, which serves as the fundamental scaling unit for modern LLM inference. While kernel-level optimizations and fine-grained overlap of communication and computation can improve absolute efficiency, the per-replica constraints imposed by KV-cache capacity, memory bandwidth, and scheduler dynamics under long-context reasoning workloads remain the same. Larger deployments primarily scale throughput via DP replication of this node-level behavior, with TP confined to the NVLink domain. Accordingly, we abstract away micro-architectural kernel specialization to quantify when system-level effects dominate and common scaling heuristics break down.

\subsection{Dataset Characterization}

\begin{figure*}[!t]
    \centering
    \begin{subfigure}[h!]{0.325\textwidth}
        \centering
        \includegraphics[width=\linewidth]{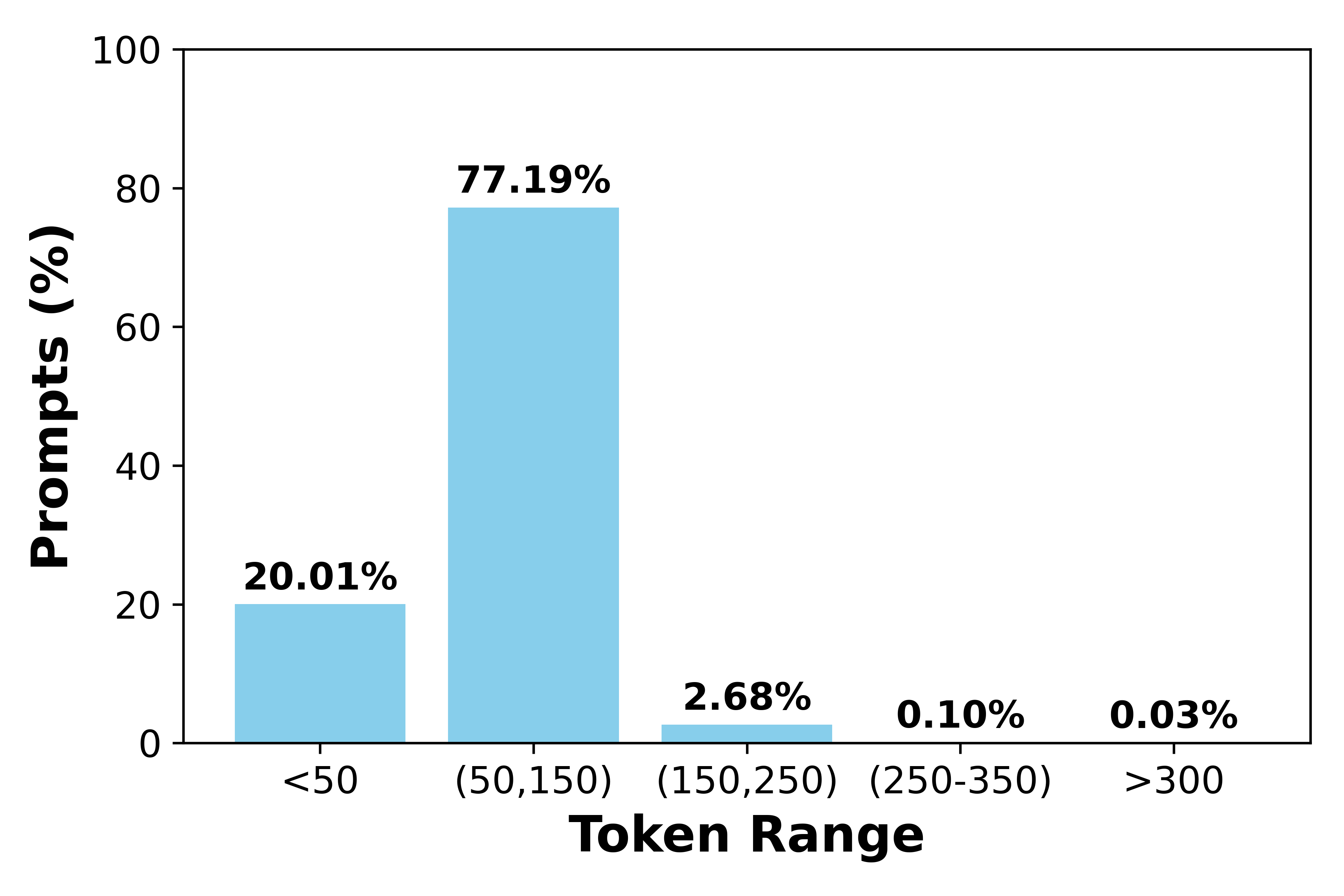}
        \caption{ISL distribution.}
        \label{fig:isl}
    \end{subfigure}
    \hfill
    \begin{subfigure}[h!]{0.33\textwidth}
        \centering
        \includegraphics[width=\linewidth]{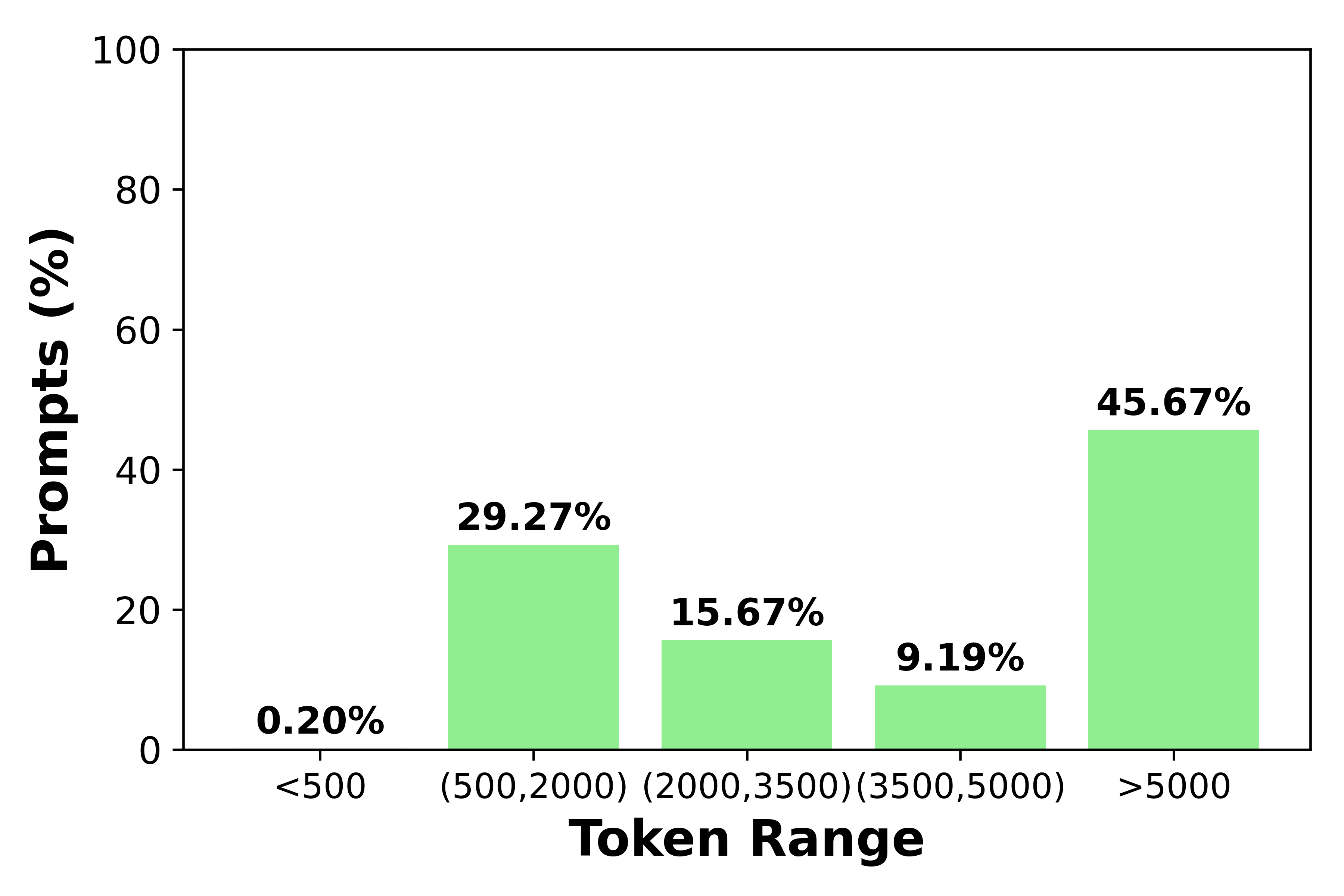}
        \caption{OSL distribution.}
        \label{fig:osl}
    \end{subfigure}
    \hfill
    \begin{subfigure}[h!]{0.325\textwidth}
        \centering
        \includegraphics[width=\linewidth]{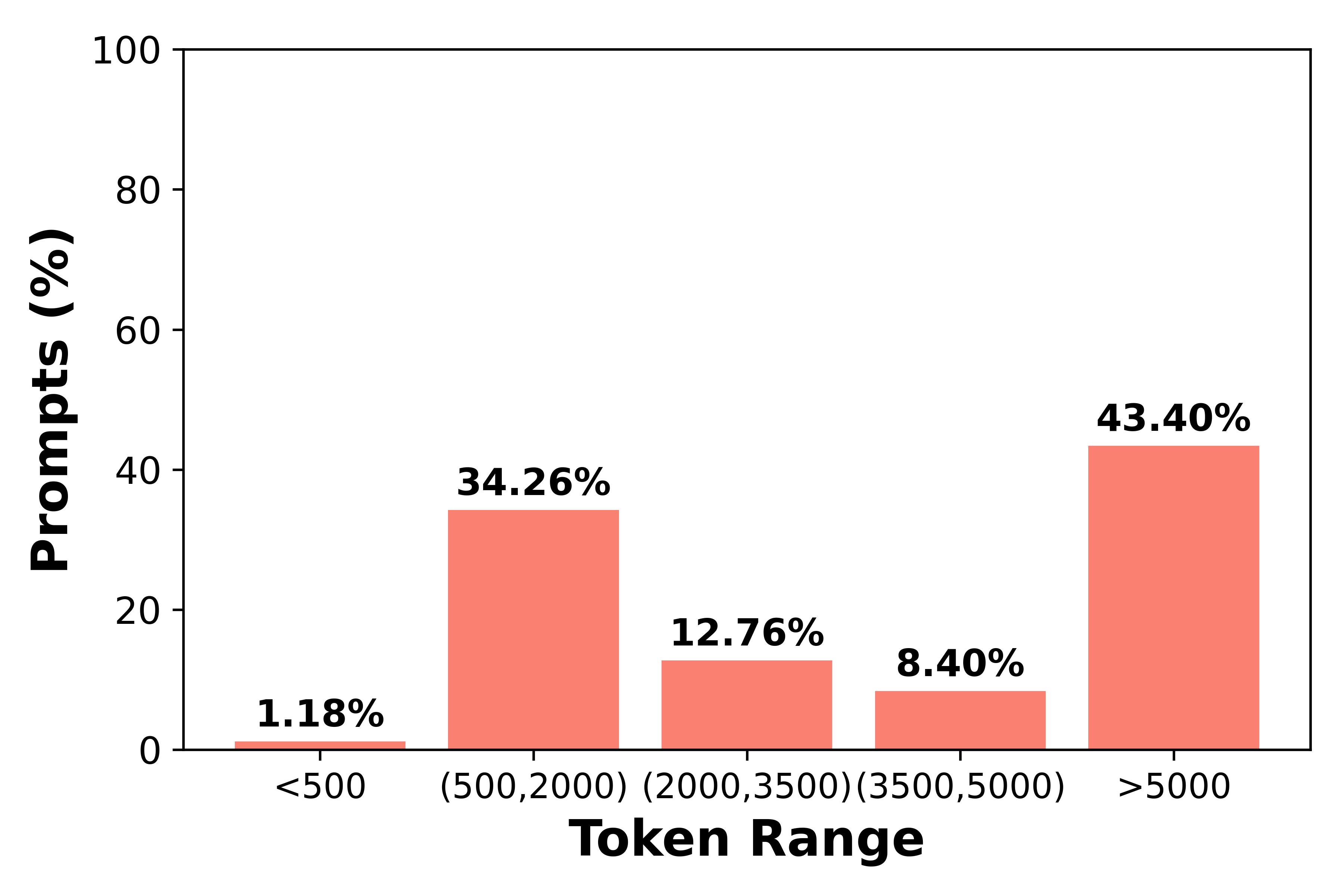}
        \caption{Reasoning token distribution.}
        \label{fig:reason}
    \end{subfigure}
     \caption{Input, output, and reasoning token distributions for 100k samples from Meta's Natural Reasoning dataset.}
    \label{fig:dataset-100ksamples}
    \vspace{-10pt}
\end{figure*}

\begin{figure*}[!t]
    \centering
    \begin{subfigure}[h!]{0.24\textwidth}
        \centering
        \includegraphics[width=\linewidth]{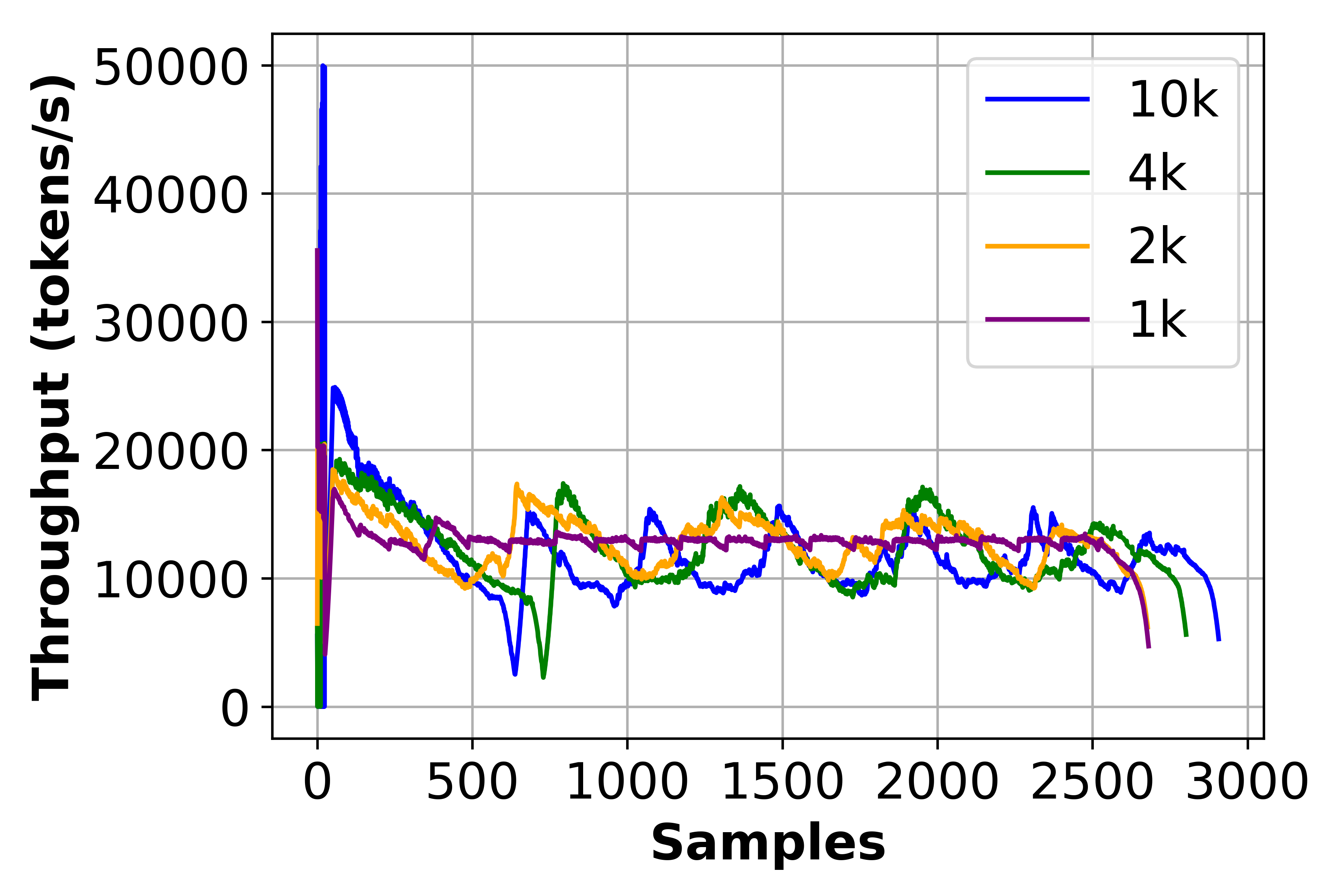}
        \caption{Generation Throughput}
        \label{fig:figure8-gen}
    \end{subfigure}
    \hfill
    \begin{subfigure}[h!]{0.24\textwidth}
        \centering
        \includegraphics[width=\linewidth]{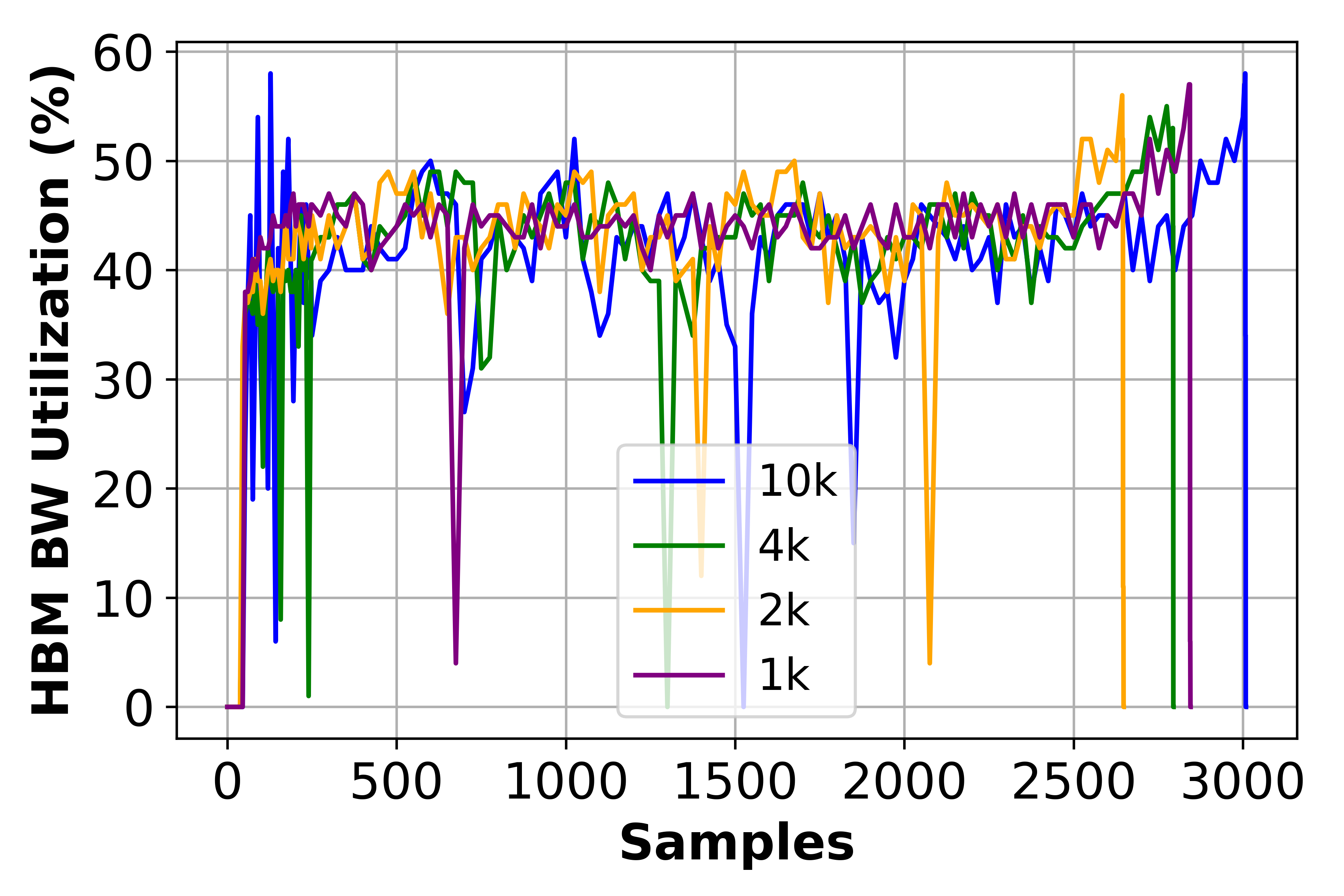}
        \caption{Average HBM BW Util.}
        \label{fig:figure8-bw}
    \end{subfigure}
    \hfill
    \begin{subfigure}[h!]{0.24\textwidth}
        \centering
        \includegraphics[width=\linewidth]{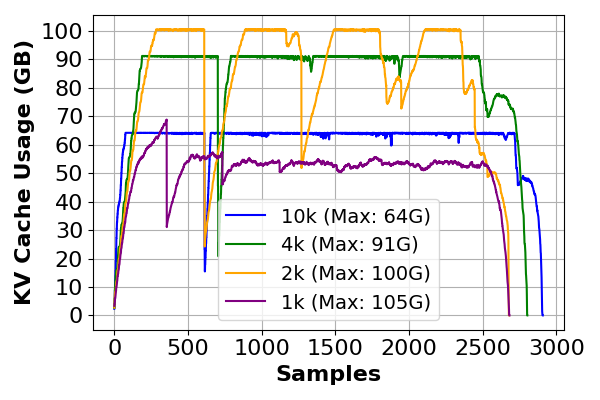}
        \caption{Aggregated KV Cache Util.}
        \label{fig:figure8-kv}
    \end{subfigure}
    \hfill
    \begin{subfigure}[h!]{0.24\textwidth}
        \centering
        \includegraphics[width=\linewidth]{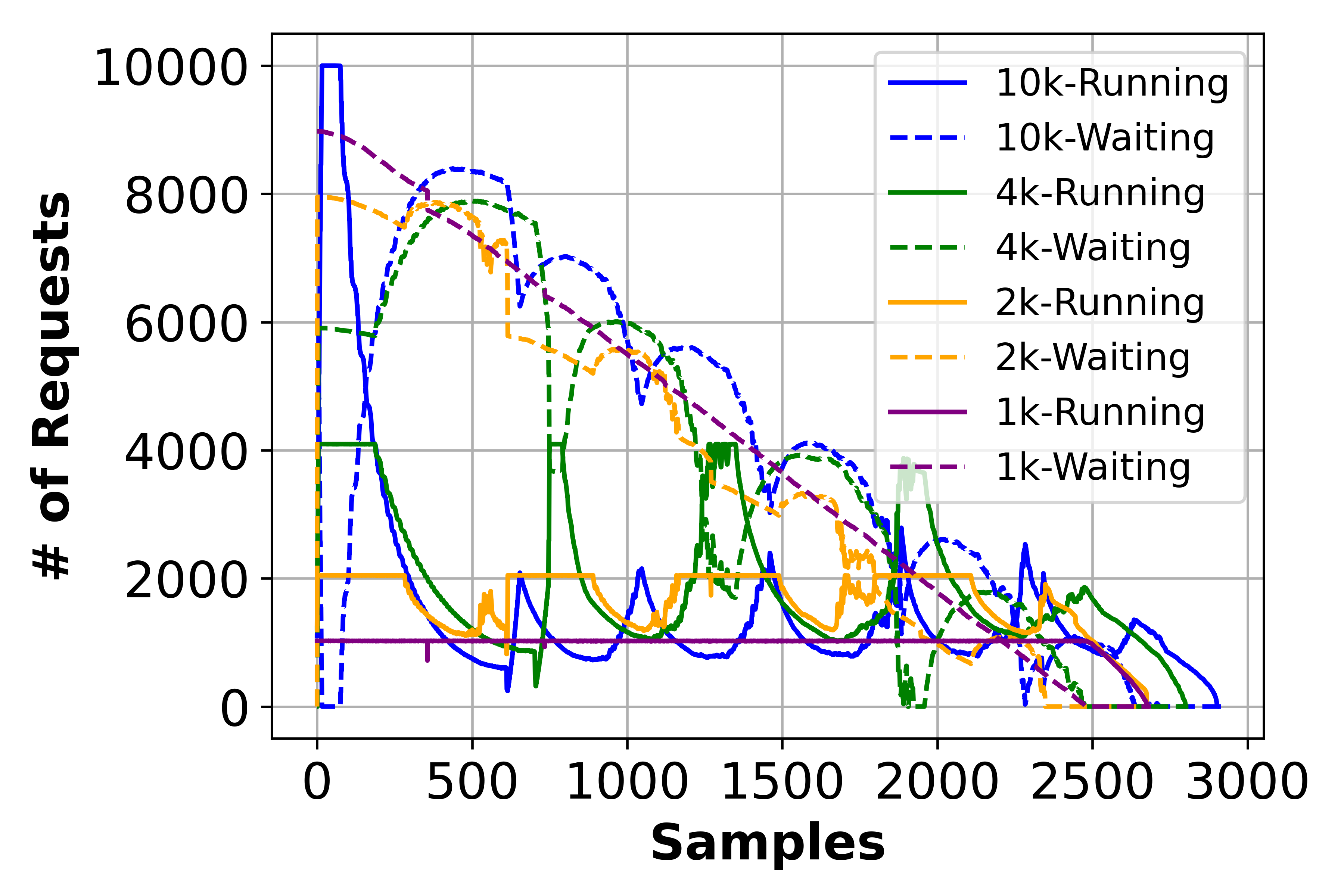}
        \caption{Request Analysis}
        \label{fig:figure8-reqs}
    \end{subfigure}
    \hfill
    \caption{Timeline of inference engine metrics on scaling the number of sequences for DeepSeek-8B on one H200 GPU.}
    \label{fig:figure8-num-seq-scaling-timeline}
    \vspace{-10pt}
\end{figure*}

To evaluate the ``Reasoning Shift'', we utilize Meta's Natural Reasoning dataset~\cite{yuan2025naturalreasoning}, comprised of 1.15 million multi-hop reasoning and commonsense inference samples. Unlike standard chat workloads dominated by input processing (prefill), this dataset forces models into a generation-heavy regime. 
Figure~\ref{fig:dataset-100ksamples} illustrates the divergent token length distributions that characterize “reasoning” versus traditional generative LLM workloads. Specifically, it compares input sequence lengths (ISL) versus output sequence lengths (OSL), highlighting that the majority of prompts remain relatively short while their outputs become exceptionally long.
A thorough analysis of 100k random samples uncovers a distinct workload profile:
\begin{itemize} [topsep=0pt,itemsep=0pt,leftmargin=12pt]
    \item \textbf{Input Sequence Length (ISL):} 77\% of prompts are 50--150 tokens, and very few exceed 300 tokens (Figure~\ref{fig:isl}).
    \item \textbf{Output Sequence Length (OSL):} 45\% of responses exceed 5000 tokens, indicating not only larger outputs but also complex reasoning chains (Figure~\ref{fig:osl}).
    \item \textbf{Reasoning Density:} Similar to OSL, reasoning steps are highly verbose, with 43.04\% of responses contain over 5000 ``reasoning tokens'' reflecting detailed multi-step reasoning token generation (Figure~\ref{fig:reason}).
\end{itemize}

Architecturally, these distributions underscore a key stress point: even modest numbers of reasoning requests can saturate GPU memory capacity due to massive KV cache footprints from long outputs, well before reaching full compute utilization. This shifts the paradigm from prefill compute density to decode memory bandwidth/capacity and interconnect latency, providing critical motivation for treating memory capacity as a first-class design parameter-- since conventional focus on FLOPs/bandwidth alone overlooks this token explosion effect inherent to reasoning workloads.

\subsection{Model Selection}
We evaluate a spectrum of models representing both the current dense architecture and the emerging reasoning-centric sparse architectures. 

\paragraph*{\bf Small-Scale Reasoning (Distilled)} We utilize DeepSeek-R1-Distill-Llama/Qwen variants (Llama-8B, Qwen-14B, Qwen-32B, Llama-70B). These models are architecturally dense but fine-tuned to output long chain-of-thought traces, allowing us to test the capacity trap of running heavy reasoning workloads on commodity-class parameters. These models utilize Grouped-Query Attention (GQA) to reduce the KV footprint by $3\times$--$8\times$ compared to the classical Multi-Head Attention (MHA) but the memory cost remains linear with layer count. For instance, the 32B model ($\approx$64 layers) consumes $\approx$262 KB/token in FP16, whereas the 70B model ($\approx$80 layers) reaches $\approx$328 KB/token.

\paragraph*{\bf Frontier-Scale Baselines}
We consider the case of Llama-3.1-405B, a standard dense transformer serving as the heavy-compute baseline. With $\approx$126 layers and dense activation, it exerts maximum pressure on HBM bandwidth. Its KV footprint is massive ($\approx$1.05 MB/token in FP16), necessitating aggressive quantization or paging for long-context inference. Next, we consider DeepSeek-R1-671B, a MoE model with $\approx$37B active parameters per token utilizing Multi-Head Latent Attention (MLA), which compresses the KV cache into a low-rank latent vector (smaller than GQA). This architectural choice decouples KV size from the number of attention heads, allowing R1 to sustain long reasoning contexts with a significantly lower memory footprint per generated token compared to a dense model of equivalent scale.

\subsection{Profiling Methodology} 
We employ a multi-layered instrumentation strategy, capturing high-level service metrics via the inference engine and low-level resource telemetry via hardware counters. This dual-view approach allows us to correlate end-to-end latency artifacts with specific micro-architectural bottlenecks.

\begin{itemize}[topsep=0pt,itemsep=0pt,leftmargin=12pt] 
    \item \textbf{Time-To-First-Token (TTFT):} The latency from request arrival to the generation of the first token. In reasoning workloads, TTFT is dominated by the prefill phase and queueing delays. High TTFT indicates prefill compute saturation or head-of-line blocking~\cite{saereesitthipitakprophet} caused by long-running decode phases of prior requests. 
    \item \textbf{Time-Per-Output-Token (TPOT):} The average inter-token latency during the generation phase. This metric is a direct proxy for memory bandwidth efficiency during autoregressive decoding. Increases in TPOT signal HBM bandwidth saturation or excessive communication overhead in tensor-parallel configurations. 
    \item \textbf{Generation Throughput:} The aggregate number of tokens generated per second across all GPUs. This system-level metric captures the efficacy of batching; sublinear scaling of throughput with batch size reveals the concurrency wall where memory capacity limits active slots. 
    \item \textbf{End-to-End (E2E) Latency:} The total elapsed time from request submission to the completion of the final token. This metric aggregates queueing delays, prefill compute, and the extended decoding phase.
    \item \textbf{Request Lifecycle Tracking:} We trace the state transitions of individual requests to decompose latency into Waiting (queueing) and Running (execution) components. This granular tracking allows us to isolate scheduler-induced delays, such as preemption or admission throttling caused by KV-cache fragmentation, from hardware execution time.
    \item \textbf{GPU/HBM Bandwidth Utilization:} Measured via nvidia-smi to track GPU utilization and memory bandwidth. 
    \item \textbf{KV-Cache Saturation:} We monitor the KV cache utilization against the total allocated KV cache. This metric is the critical indicator of the capacity trap; nearing 100\% saturation forces the scheduler to preempt requests to free memory, causing catastrophic spikes in end-to-end latency due to re-computation costs. 
\end{itemize}

Together, these metrics allow us to distinguish compute saturation, memory bandwidth saturation, and KV-capacity-driven preemption effects, enabling causal attribution of end-to-end latency degradation under scaling.

\section{Analysis I: Capacity Trap for Small Models}
\label{sec:analysis_small}

This section characterizes the performance boundaries of small-sized models (8B--32B) when serving memory-intensive reasoning workloads. We investigate the ``What-If'' scenario: \textit{Does maximizing GPU occupancy via high concurrency yield sustainable throughput for reasoning, or does it trigger a resource collapse?}

\subsection{The Concurrency-Capacity Trade-off}
\label{sec:concurrency_tradeoff}

\begin{figure*}[!t]
    \centering
    \begin{subfigure}[h!]{0.24\textwidth}
        \centering
        \includegraphics[width=\linewidth]{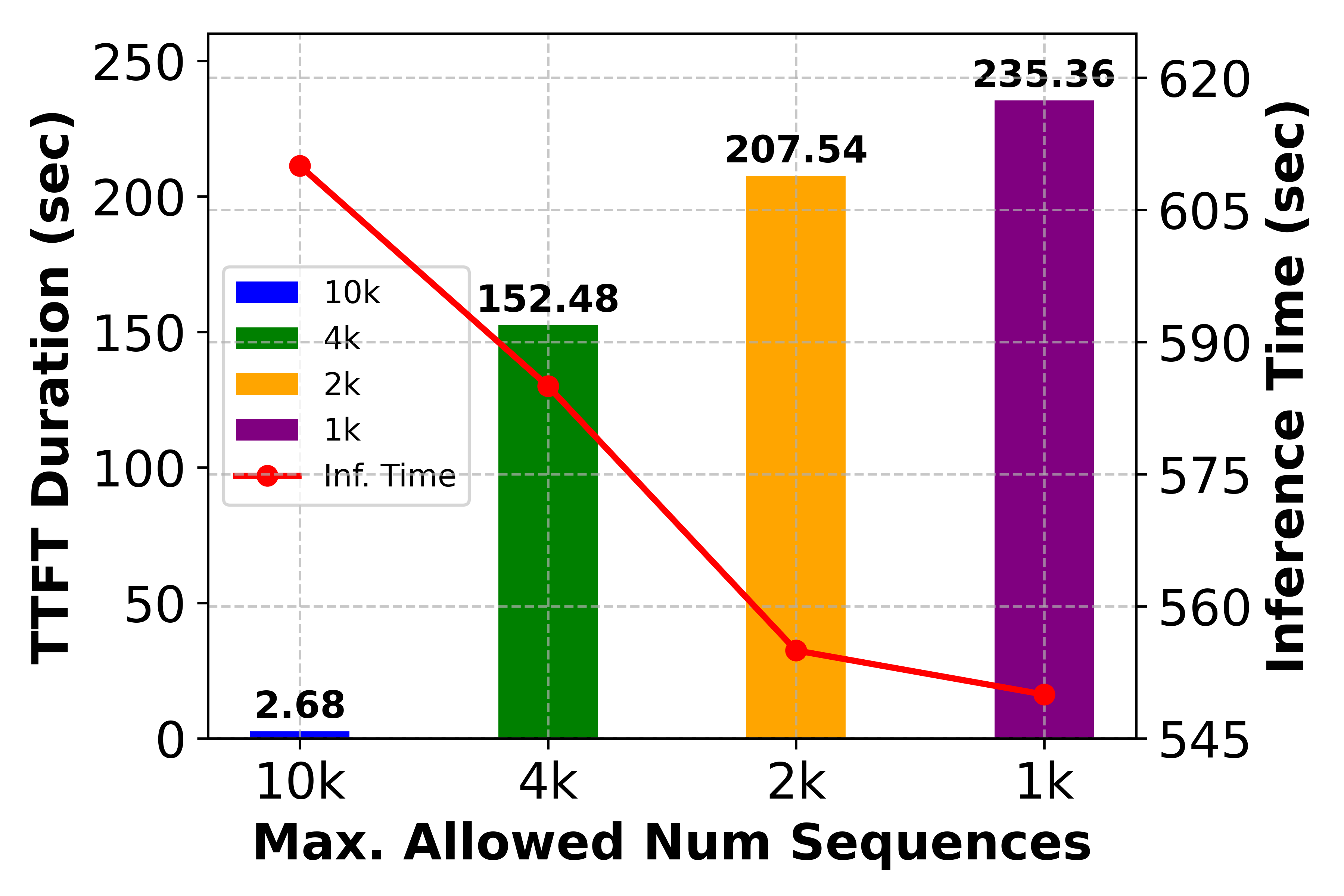}
        \caption{Average Request TTFT and Inference Duration}
        \label{fig:figure8-ttft}
    \end{subfigure}
    \hfill
    \begin{subfigure}[h!]{0.24\textwidth}
        \centering
        \includegraphics[width=\linewidth]{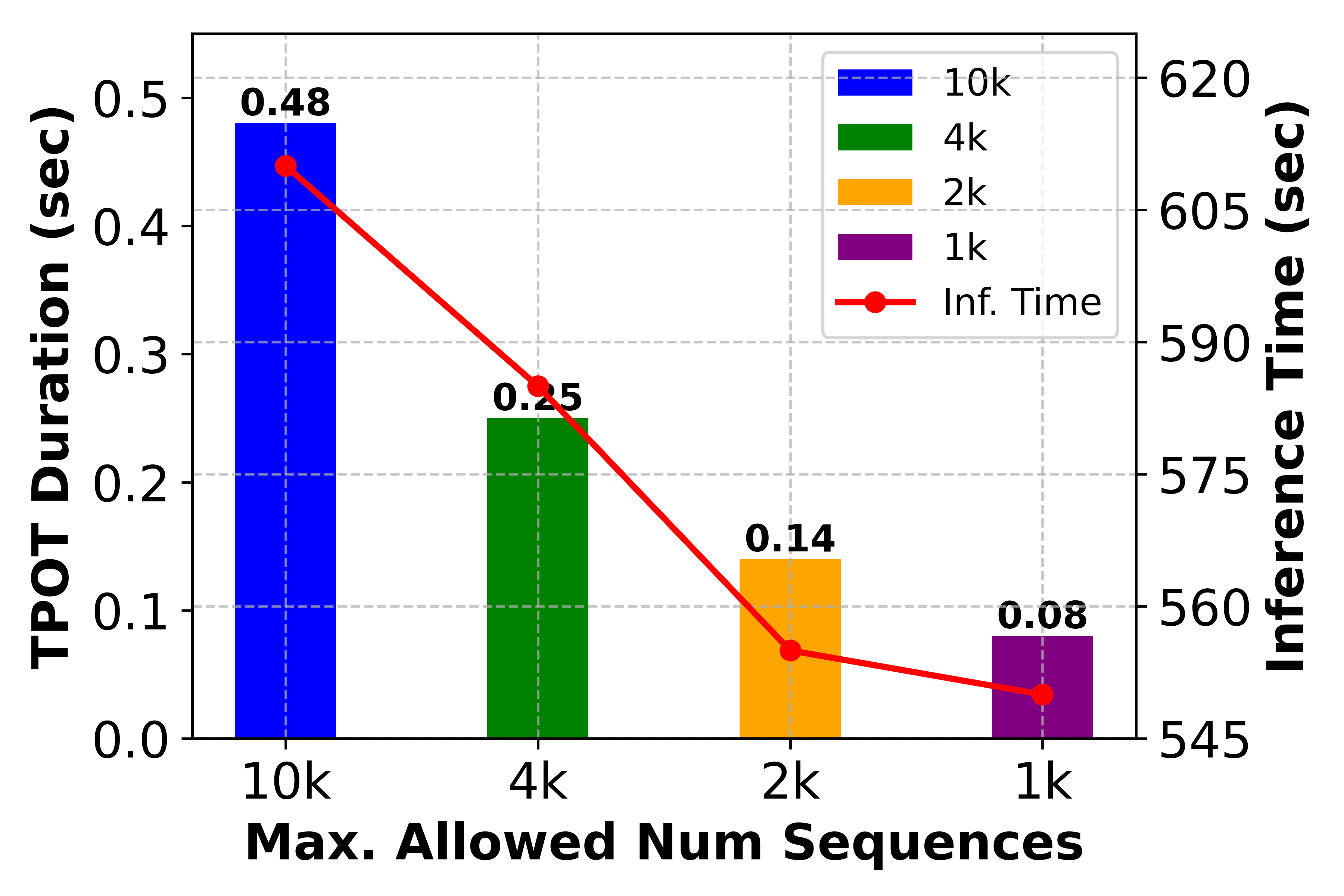}
        \caption{Average Request TPOT and Inference Duration}
        \label{fig:figure8-tpot}
    \end{subfigure}
    \hfill
    \begin{subfigure}[h!]{0.24\textwidth}
        \centering
        \includegraphics[width=\linewidth]{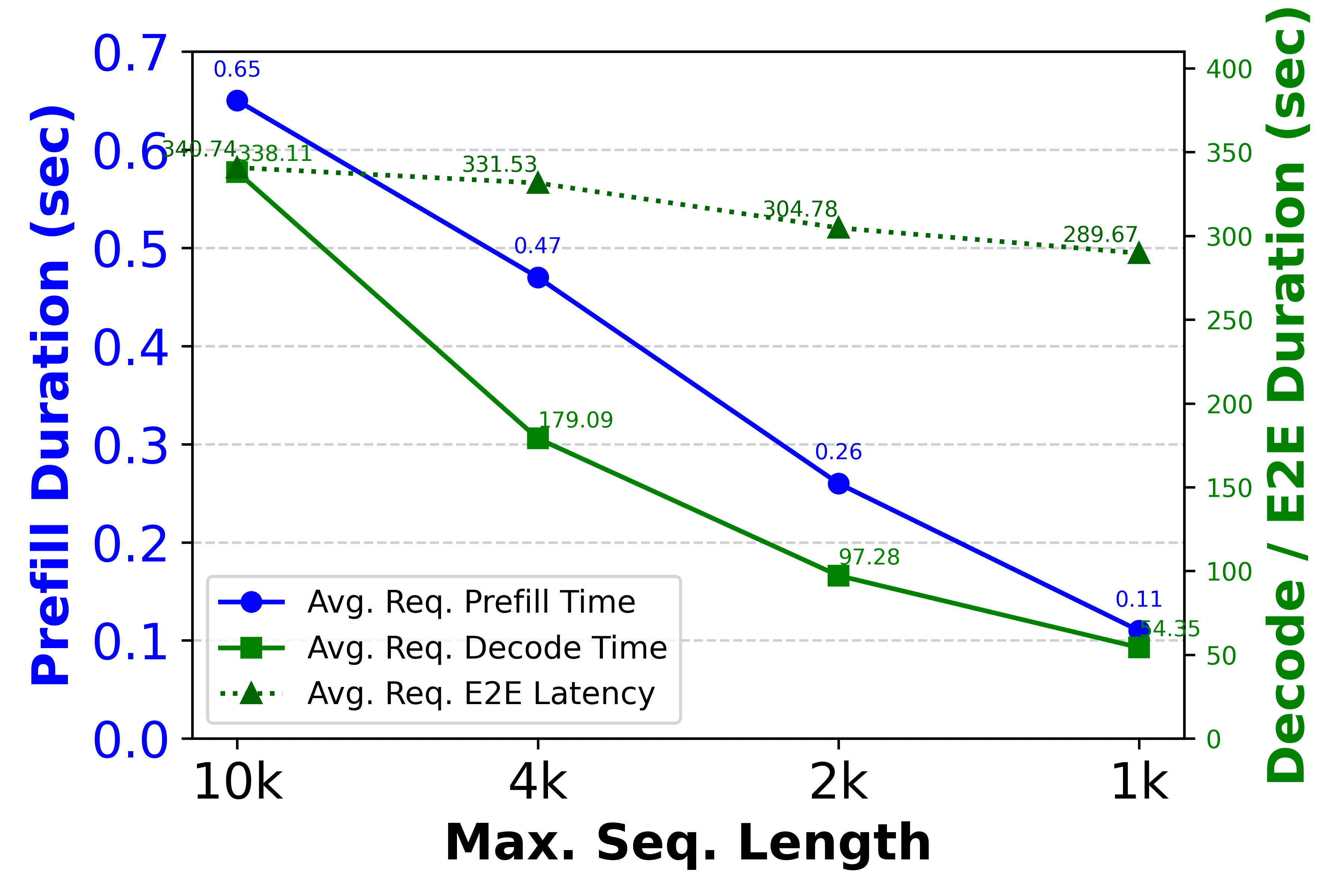}
        \caption{Average Request Prefill, Decode, and E2E latency.}
        \label{fig:figure8-prefill-decode-e2e}
    \end{subfigure}
    \hfill
    \begin{subfigure}[h!]{0.24\textwidth}
        \centering
        \includegraphics[width=\linewidth]{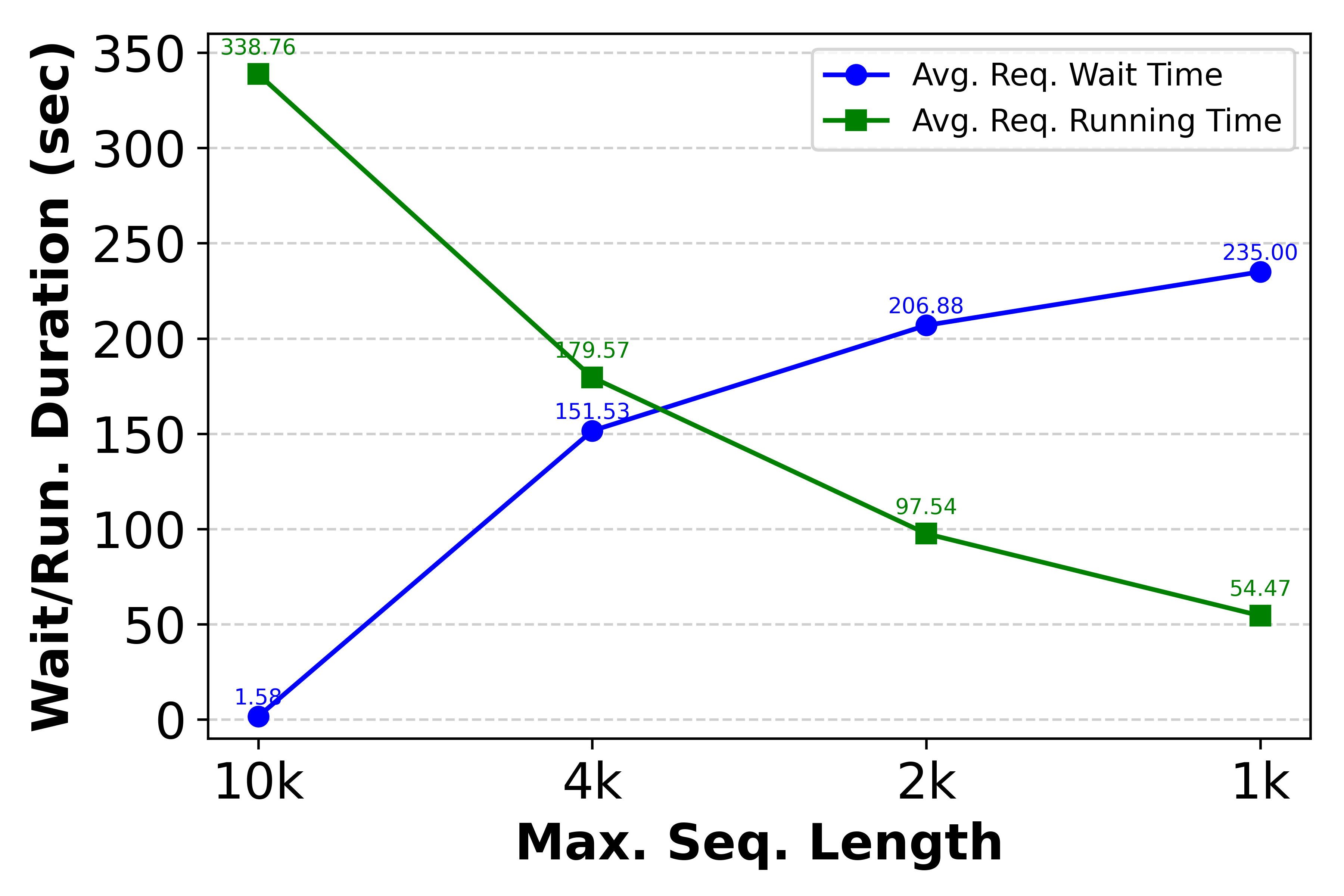}
        \caption{Average Request Waiting and Running Durations.}
        \label{fig:figure8-queue-inf}
    \end{subfigure}
    \caption{Overall serving statistics on scaling maximum number of sequences for DeepSeek-8B on one H200 GPU.}
    \label{fig:figure8-num-seq-scaling-overall}
    \vspace{-10pt}
\end{figure*}

\begin{figure*}[!t]
    \centering
    \begin{subfigure}[h!]{0.24\textwidth}
        \centering
        \includegraphics[width=\linewidth]{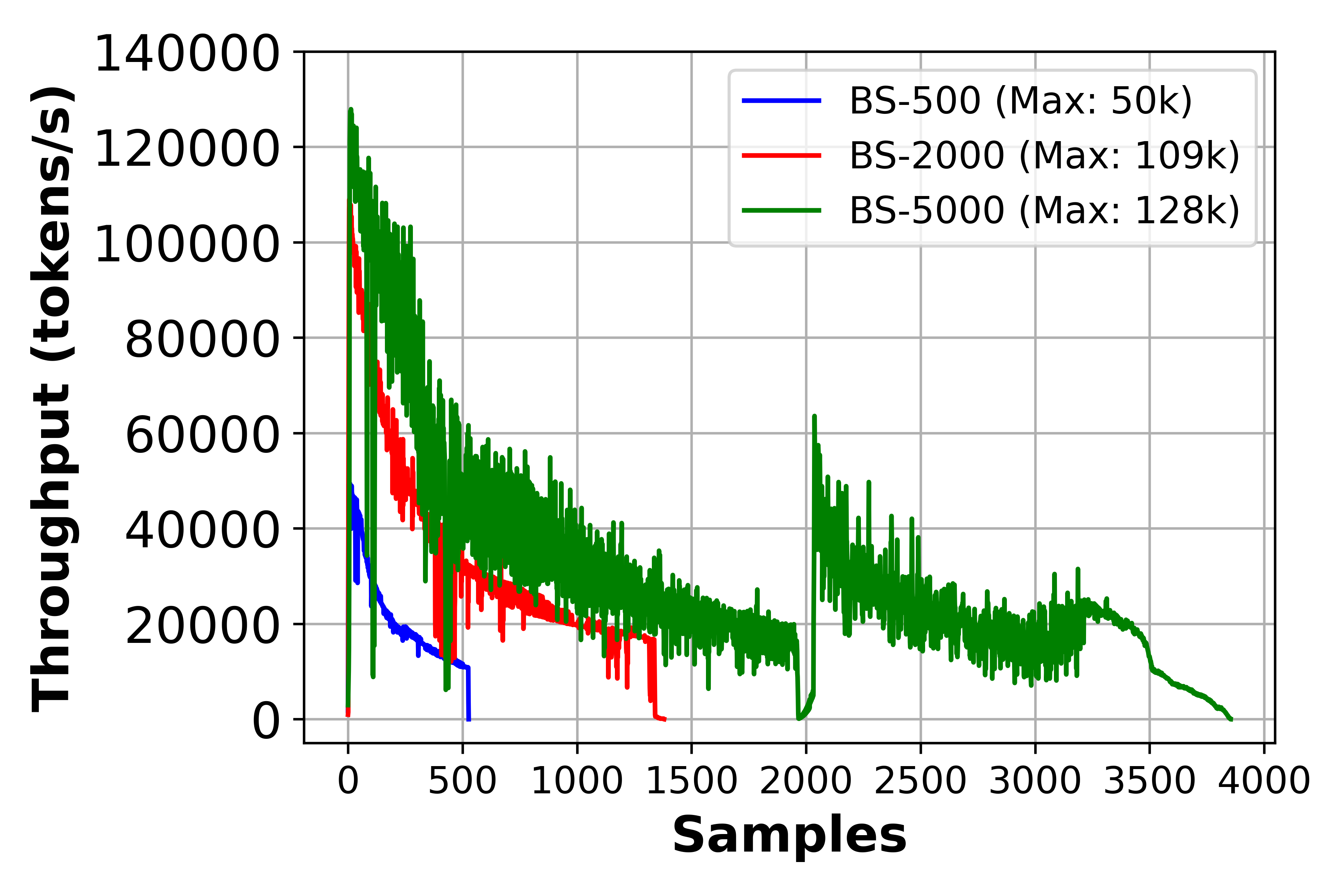}
        \caption{Generation Throughput}
        \label{fig:figure2-gen}
    \end{subfigure}
    \hfill
    \begin{subfigure}[h!]{0.24\textwidth}
        \centering
        \includegraphics[width=\linewidth]{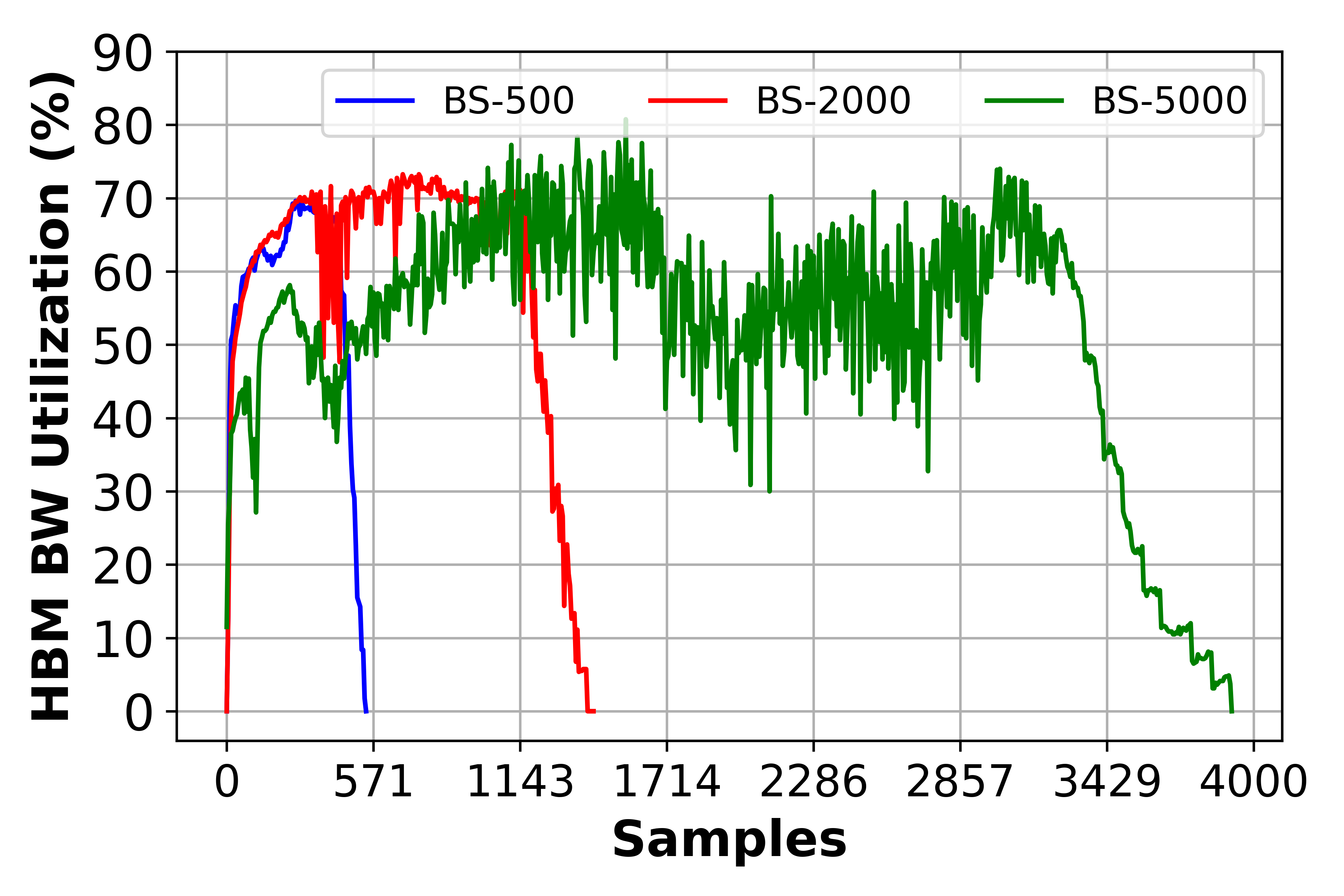}
        \caption{Average HBM BW Util.}
        \label{fig:figure2-bw}
    \end{subfigure}
    \hfill
    \begin{subfigure}[h!]{0.24\textwidth}
        \centering
        \includegraphics[width=\linewidth]{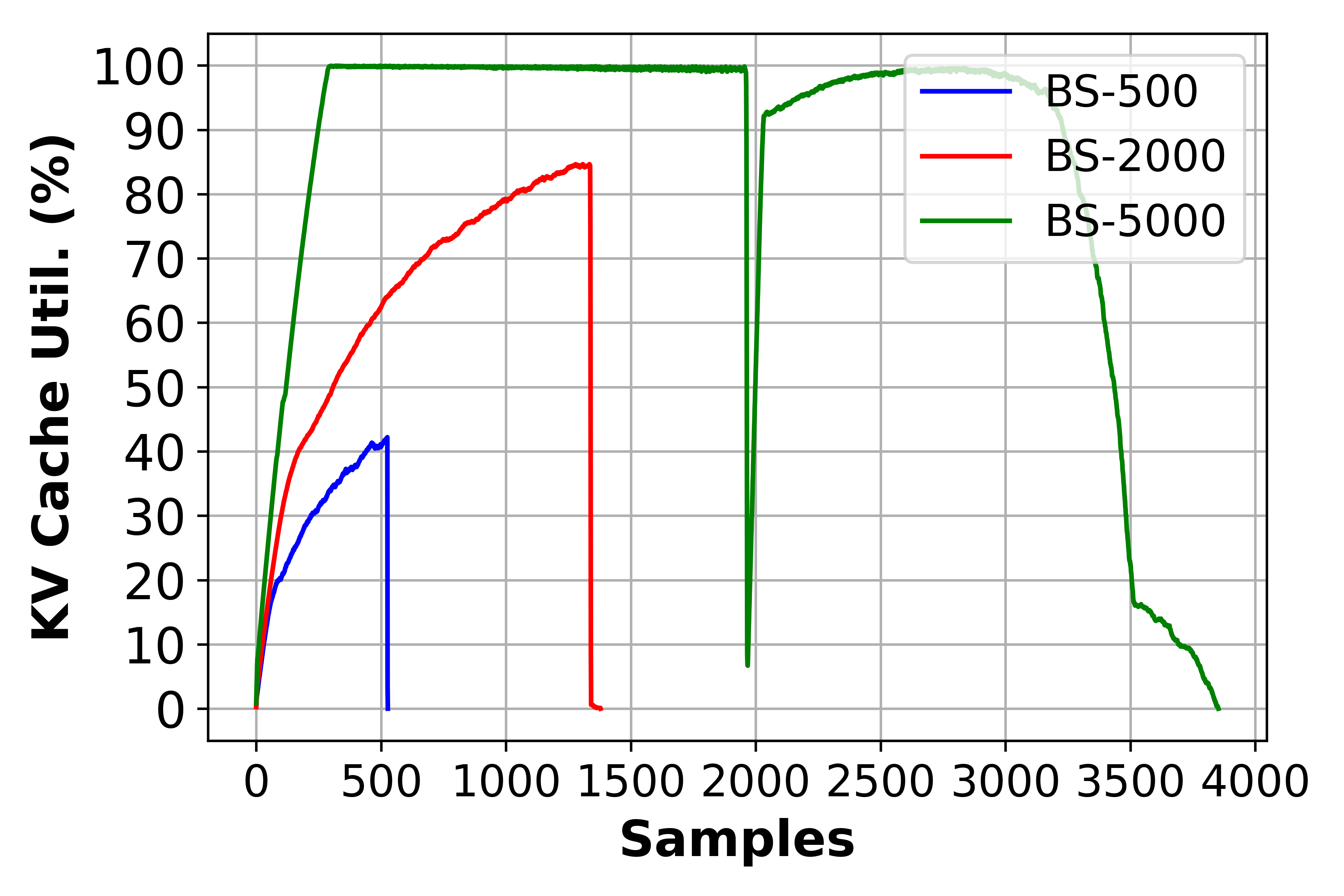}
        \caption{Aggregated KV Cache Util.}
        \label{fig:figure2-kv}
    \end{subfigure}
    \hfill
    \begin{subfigure}[h!]{0.24\textwidth}
        \centering
        \includegraphics[width=\linewidth]{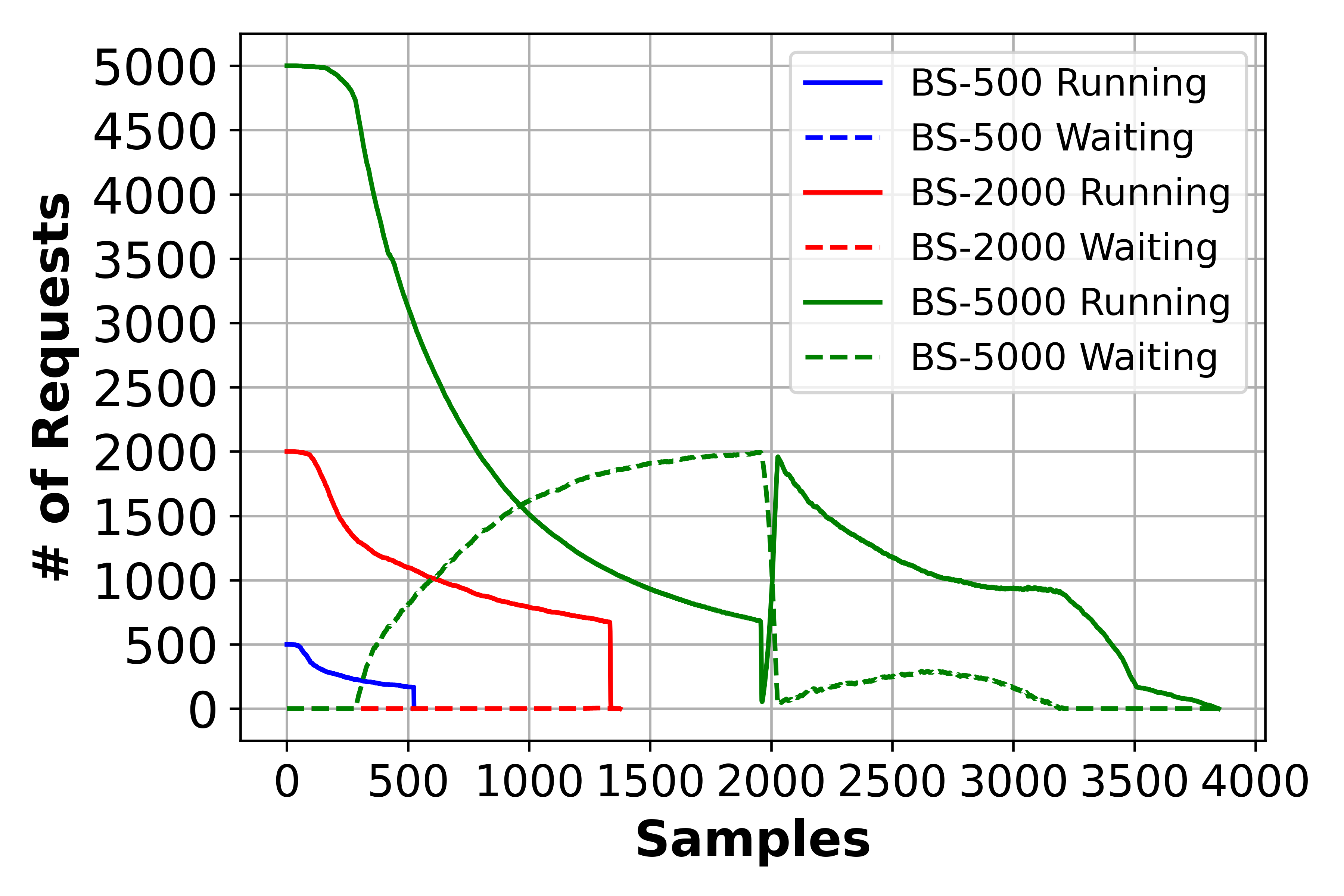}
        \caption{Request Analysis}
        \label{fig:figure2-reqs}
    \end{subfigure}
    \caption{Batch size scaling for DeepSeek-8B on 8x H200 GPUs with 8-way DP.}
    \label{fig:figure2-batch-scaling}
    \vspace{-10pt}
\end{figure*}

In traditional inference, increasing the maximum number of concurrent sequences (\texttt{max\_num\_seqs}) is the primary lever to amortize kernel launch overheads and hide HBM latency. We evaluate this heuristic with DeepSeek-8B model for a batch of 10K input sequences while scaling concurrency with `max\_num\_seqs' from 1K to 10K on one H200 GPU.

\paragraph*{\bf The Occupancy-Saturation Conflict}
Our max-sequence scaling experiments (Figure~\ref{fig:figure8-num-seq-scaling-timeline}) reveal a fundamental conflict between compute utilization and memory availability. Figure~\ref{fig:figure8-num-seq-scaling-timeline} illustrates the temporal evolution of throughput, bandwidth utilization, KV occupancy, and request state as concurrency scales. While higher concurrency initially increases throughput (Figure~\ref{fig:figure8-gen}), the corresponding rise in KV utilization (Figure~\ref{fig:figure8-kv}) rapidly saturates HBM, triggering scheduler preemption events (Figure~\ref{fig:figure8-reqs}), which in turn cause drops in bandwidth utilization ((Figure~\ref{fig:figure8-bw})). The x-axis samples represent uniformly spaced measurements and are proportional to elapsed time. Increasing concurrent sequences to 10K (Figure~\ref{fig:figure8-gen}) achieves the highest \textit{initial} throughput because the scheduler can immediately saturate the SMs with prefill tokens. However, this advantage is transient due to limited memory capacity. Similarly, the bandwidth utilization (Figure~\ref{fig:figure8-bw}) closely follows the generation throughput trend, while dropping during memory saturation and re-scheduling. 

\paragraph*{\bf The Preemption Cliff} As shown in Figure~\ref{fig:figure8-kv}, the 10K configuration drives the aggregated KV usage to 100\% almost instantly. Unlike smaller batch settings (1K), which maintain a stable memory footprint, the 10K setting forces the vLLM scheduler into a thrashing regime. To prevent OOM errors, the scheduler must preempt active (Running) requests (Figure~\ref{fig:figure8-reqs}), demoting them to the Waiting queue.

\paragraph*{\bf The Re-computation Penalty} This thrashing introduces a severe latency penalty. When preempted requests are rescheduled, the engine attempts to recover their state via \textit{prefix caching}. However, under memory exhaustion, typically, the prefix match fails, and the system falls back to full prefill re-computation~\cite{KVCacheProfile-IPDPS25}. Even with successful partial matching, the overhead of searching blocks (either GPU-resident cached by vLLM or offloaded prefixes cached through LMCache~\cite{cheng2025lmcache}, MoonCake~\cite{qin2024mooncake,park2025survey}, etc.) and partial recomputations destroys tail latency stability.

\paragraph*{\bf Analytical KV Sizing} While KV-cache capacity can be analytically estimated from model and batching parameters to guide admission control, our results show that such estimates are insufficient to prevent preemptions under dynamic, long-context reasoning workloads, where scheduling effects and KV fragmentation create transient capacity spikes.

\begin{observationbox}
\label{observation-1}
\textbf{Observation 1:} 
For reasoning workloads, increasing concurrency improves GPU occupancy only until the cumulative KV footprint from long-lived reasoning tokens saturates HBM creating a Capacity Trap. Beyond this point, additional requests trigger preemption and recomputation, causing throughput gains to collapse. Hence, schedulers should avoid maximizing concurrency blindly; instead, enforce a KV-aware concurrency cap based on available HBM headroom, active sequence length, and expected decode growth.
\end{observationbox}

\subsection{The Latency Decoupling: TTFT vs. TPOT}
\label{sec:latency_decoupling}
While the Capacity Trap explains \textit{why} the system struggles, the impact on service quality is best understood by studying the request TTFT, TPOT, and E2E. Figure~\ref{fig:figure8-num-seq-scaling-overall} quantifies this Pareto frontier. Importantly, the convex E2E latency curve reflects not a tuning artifact, but a fundamental trade-off between admission latency and KV-limited decode progress, defining a concurrency sweet spot intrinsic to reasoning workloads. We define inference time as the total execution time of prefill and decode after request scheduling, while the E2E latency captures the full request lifecycle, including queuing, scheduling, and inference.

\begin{figure}[ht!]
    \centering
    \includegraphics[width=0.95\linewidth]{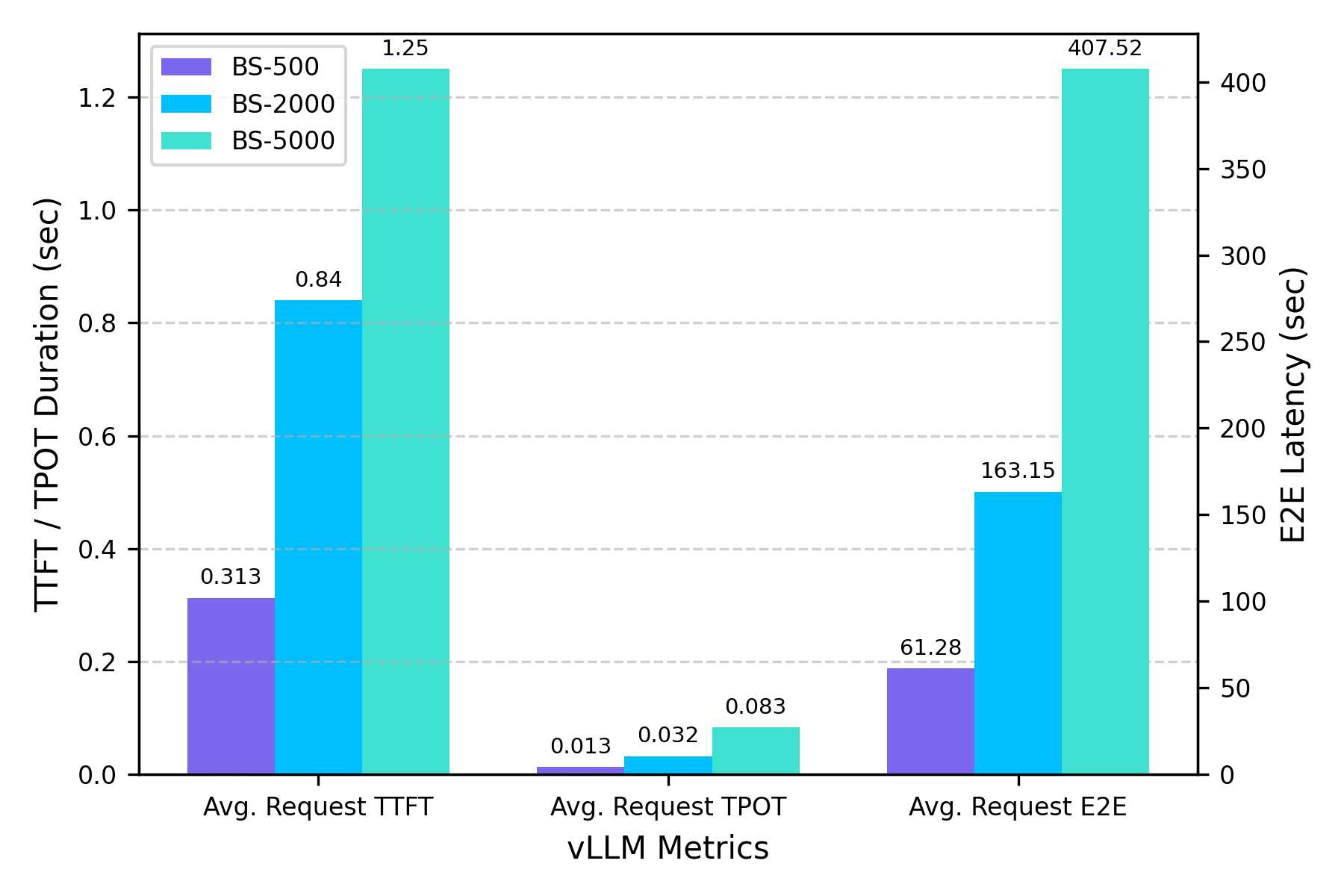}
    \caption{vLLM Metrics for 500, 2000 and 5000 batch sizes for DS-8B for 8-way DP.}
    \label{fig:figure-2-ttft-tpot-e2e}
    \vspace{-10pt}
\end{figure}

\paragraph*{\bf The Inverse Scaling Law}
We observe a sharp divergence in metric behavior as concurrency scales: 
(1) \textbf{TTFT (Queue-Bound):} TTFT is minimized at maximum concurrency (10K), dropping to $\approx$2.7s (Figure~\ref{fig:figure8-ttft}). This is intuitive: with more open slots, a request spends less time in the admission queue. High concurrency more efficiently utilizes GPU's parallelization capabilities;
(2) \textbf{TPOT (Bandwidth and Capacity Bound):} Conversely, TPOT degrades linearly with concurrency, rising from $\approx$0.08s at 1K to $\approx$0.48s at 10K (Figure~\ref{fig:figure8-tpot}). In addition to being bandwidth-bound (frequent reads and limited writes per iteration), the decode phase is also capacity bound. For an increasing number of requests, preemptions due to memory constraints worsen TPOT.

\paragraph*{\bf The End-to-End Convexity}
The End-to-End~(E2E) latency (Figure~\ref{fig:figure8-prefill-decode-e2e}) reveals the net effect of these opposing forces and exhibits a non-monotonic convexity with a distinct sweet spot at the concurrency level of $\approx$2K sequences. This 2K point provides the best latency--capacity trade-off as it captures most of the latency improvement from reducing concurrency from 10K to 1K, while still preserving twice the sequence capacity of the 1K setting. Reducing further to 1K yields diminishing returns relative to the loss in context capacity, as reflected by higher TTFT (Figure~\ref{fig:figure8-ttft}) and increased request wait time (Figure~\ref{fig:figure8-queue-inf}), despite only marginal capacity gains. In contrast, beyond 2K, capacity degrades rapidly due to increasing decode KV pressure (Figure~\ref{fig:figure8-ttft}--~\ref{fig:figure8-queue-inf}). We also agree that the observed batch-E2E behavior is sub-linear and will correct this characterization.

\begin{itemize}[topsep=0pt,itemsep=0pt,leftmargin=12pt] 
    \item \textbf{Low Concurrency Regime ($<$2K):} E2E latency is dominated by \textit{Queueing Delay}. Although the GPU processes active requests fast (low TPOT), new requests wait too long to enter the system (Figure~\ref{fig:figure8-queue-inf}) indicating insufficient in-flight work to keep the system saturated, resulting in poor throughput efficiency despite low per-request latency.
    \item \textbf{High Concurrency Regime ($>$2K):} E2E latency is dominated by \textit{Execution Dilution} and \textit{Preemption}. The active requests run slowly (high TPOT) due to capacity and bandwidth contention, and the tail latency spikes due to the preemption artifacts described in Observation 1. New requests enter the system the fastest but spend longer durations in the running state (Figure~\ref{fig:figure8-queue-inf}).
\end{itemize}

\begin{observationbox}
\textbf{Observation 2:} 
Reasoning workloads expose a direct trade-off between admission latency~(TTFT) and generation latency~(TPOT). Larger batches reduce queuing delay and improve TTFT, but they also reduce per-request available memory capacity and bandwidth, increasing TPOT during long decode phases. The optimal operating point is therefore the batch size where TTFT reduction no longer compensates for TPOT degradation. This motivates online batch-size tuning using TTFT, TPOT, KV occupancy, and HBM bandwidth as feedback signals.
\end{observationbox}

\begin{figure*}[!ht]
    \centering
    \begin{subfigure}[b]{0.24\textwidth}
        \centering
        \includegraphics[width=\linewidth]{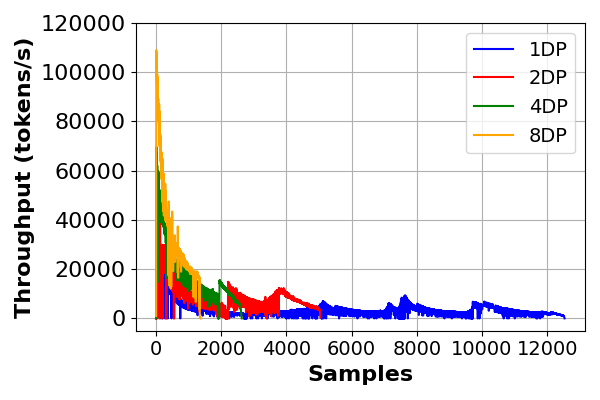}
        \caption{Generation Throughput}
        \label{fig:figure7-gen}
    \end{subfigure}
    \hfill
    \begin{subfigure}[b]{0.24\textwidth}
        \centering
        \includegraphics[width=\linewidth]{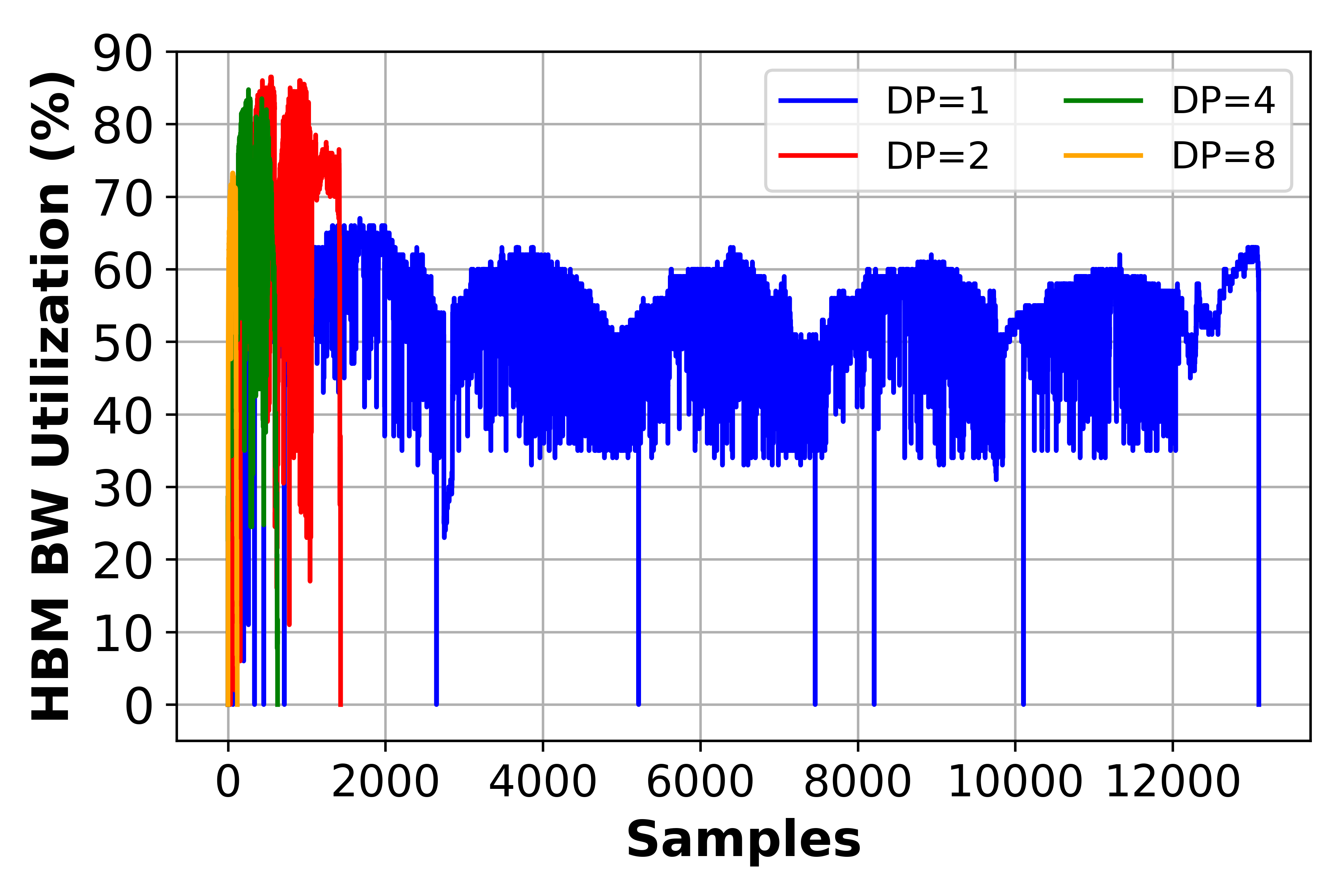}
        \caption{Average HBM BW Util.}
        \label{fig:figure7-bw}
    \end{subfigure}
    \hfill
    \begin{subfigure}[b]{0.24\textwidth}
        \centering
        \includegraphics[width=\linewidth]{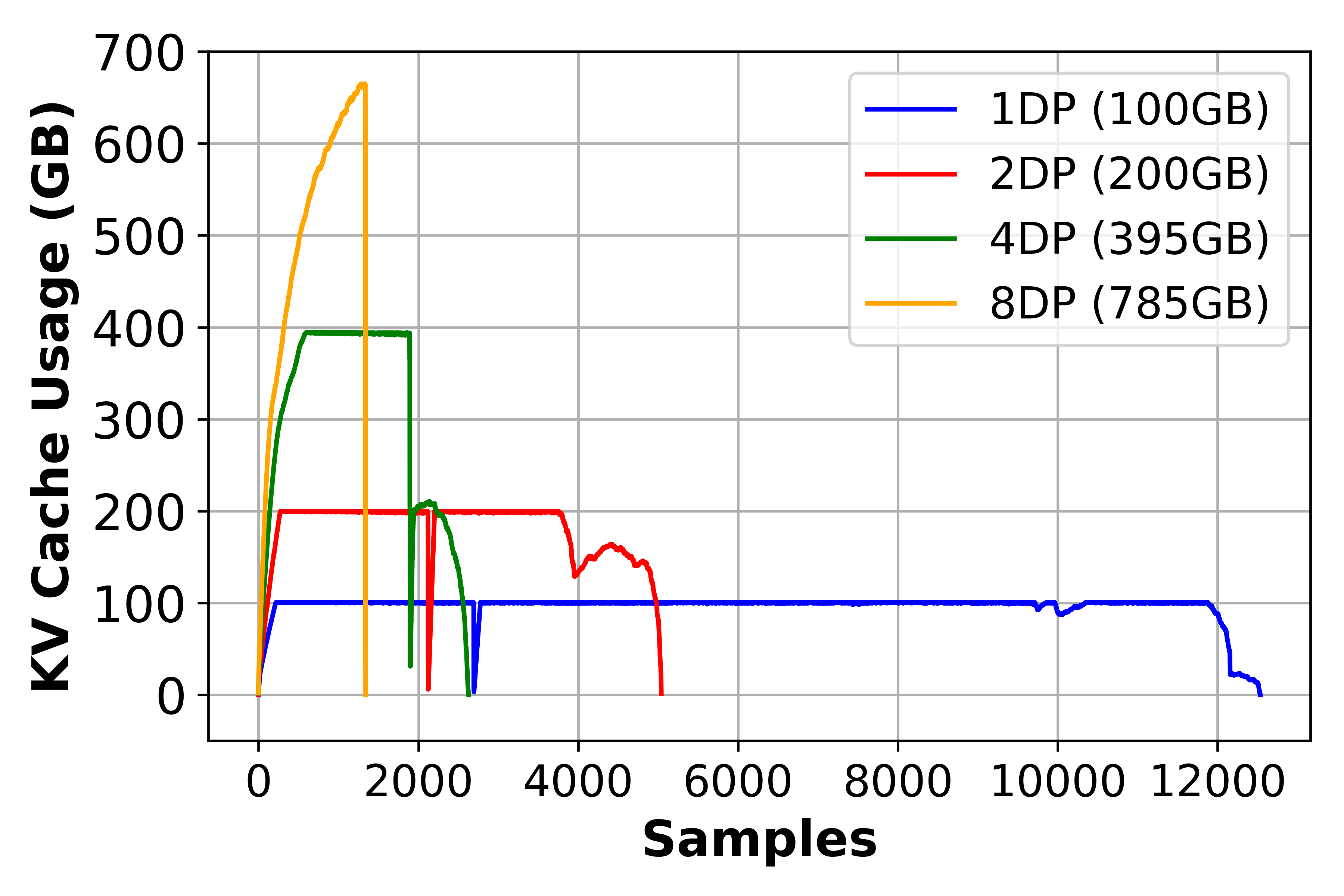}
        \caption{Aggregated KV Cache Util.}
        \label{fig:figure7-kv}
    \end{subfigure}
    \hfill
    \begin{subfigure}[b]{0.24\textwidth}
        \centering
        \includegraphics[width=\linewidth]{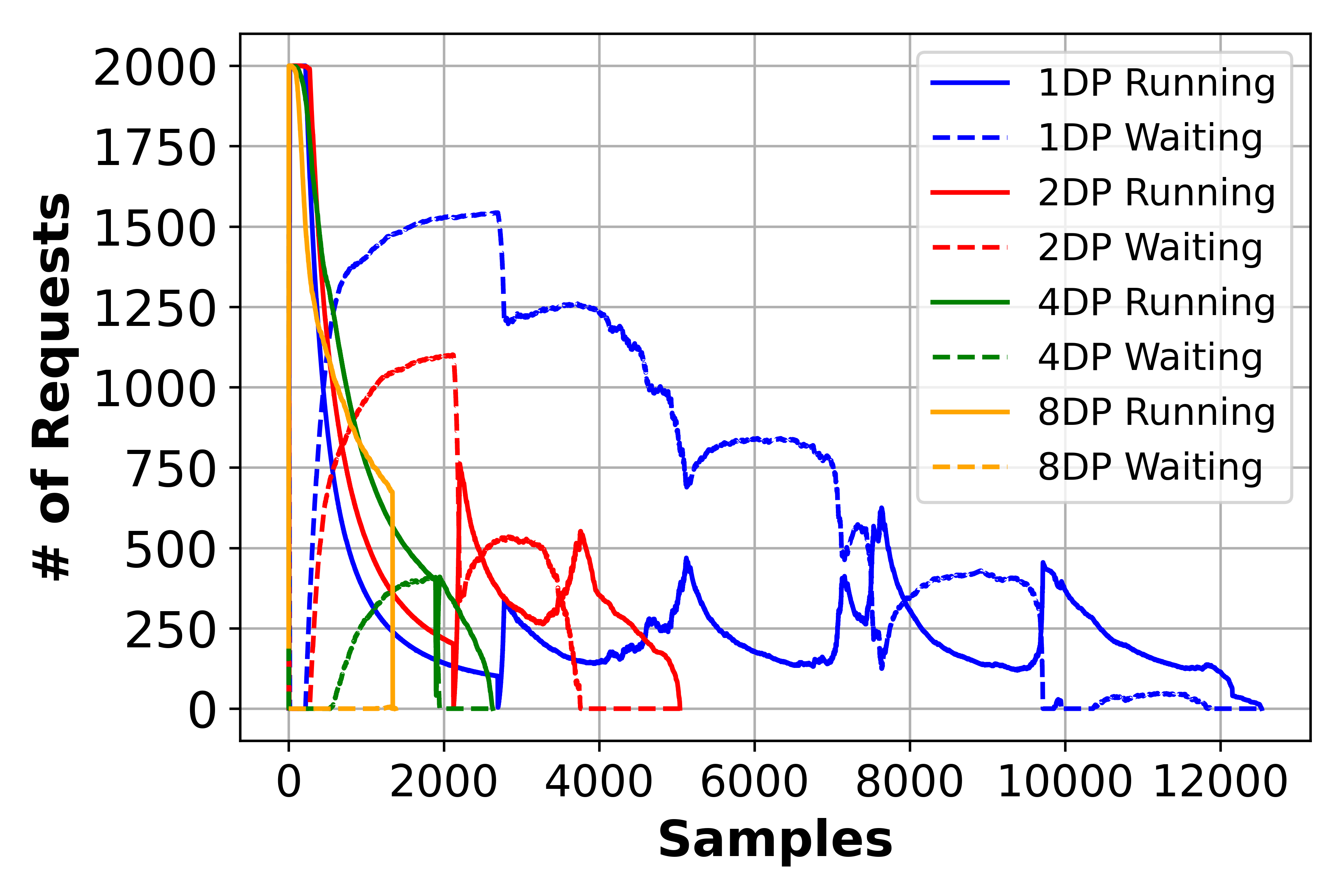}
        \caption{Request Analysis}
        \label{fig:figure7-reqs}
    \end{subfigure}
    \caption{Scale up for DeepSeek-8B model (best strategy: DP scaling).}
    \label{fig:figure7-8b-scaling}
    \vspace{-10pt}
\end{figure*}

\subsection{Batch Size Scaling on 8 GPUs}
\label{sec:batch_scaling}
To counter the effects reduced memory capacity and bandwidth during the decode phase, we use DP to scale the workload to 8 GPUs. Intuitively, DP would alleviate the pressure by distributing the requests across independent model replicas. We investigate this by fixing the cluster size (8$\times$H200, DP=8) and scaling the input \textbf{Batch Size (BS)} from 500 to 5000 requests (Figure~\ref{fig:figure2-batch-scaling}).

\paragraph*{\bf Persisting DP Memory Saturation}
Despite the cluster having $\approx$1.1 TB of aggregate HBM, Figure~\ref{fig:figure2-kv} shows that the ``Capacity Trap'' persists at scale (e.g., BS-5000) while lower load shows under-utilized KV cache. Specifically:
(1) \textbf{Throughput vs. Latency Divergence:} Scaling from BS-500 to BS-5000 increases aggregate throughput (Figure~\ref{fig:figure2-gen}). However, Figure~\ref{fig:figure-2-ttft-tpot-e2e} shows that End-to-End (E2E) latency grows sub-linearly ($61s \rightarrow 165s$). Similarly, the TTFT and TPOT also increase with increase load on the compute cluster while observing request throttling for BS-5000 due to fully utilized KV cache (Figure~\ref{fig:figure2-reqs}); 
(2) \textbf{Why DP Fails to buffer Capacity:} At BS-5000, each of the 8 GPUs is assigned $\approx$625 requests. Since DP does not pool memory, each GPU effectively operates as an isolated island facing the same overload scenario characterized in Observation 1. The HBM bandwidth utilization (Figure~\ref{fig:figure2-bw}) saturates but exhibits dips corresponding to scheduler thrashing, proving that adding GPUs via DP scales \textit{compute} but does not resolve the per-stream \textit{memory pressure}, which we investigate next.

\begin{observationbox}
\textbf{Observation 3:} 
Data Parallelism allows the system to admit more requests across replicas, but each GPU still stores a full copy of the model and independently faces the same KV capacity limit. Thus, DP improves aggregate serving capacity only when each replica remains below its local HBM saturation point. For reasoning-heavy workloads, DP should be combined with admission control or memory-aware routing to prevent each replica from independently entering a preemption-heavy regime.
\end{observationbox}

\begin{figure}[!ht]
    \centering
    \includegraphics[width=0.99\linewidth]{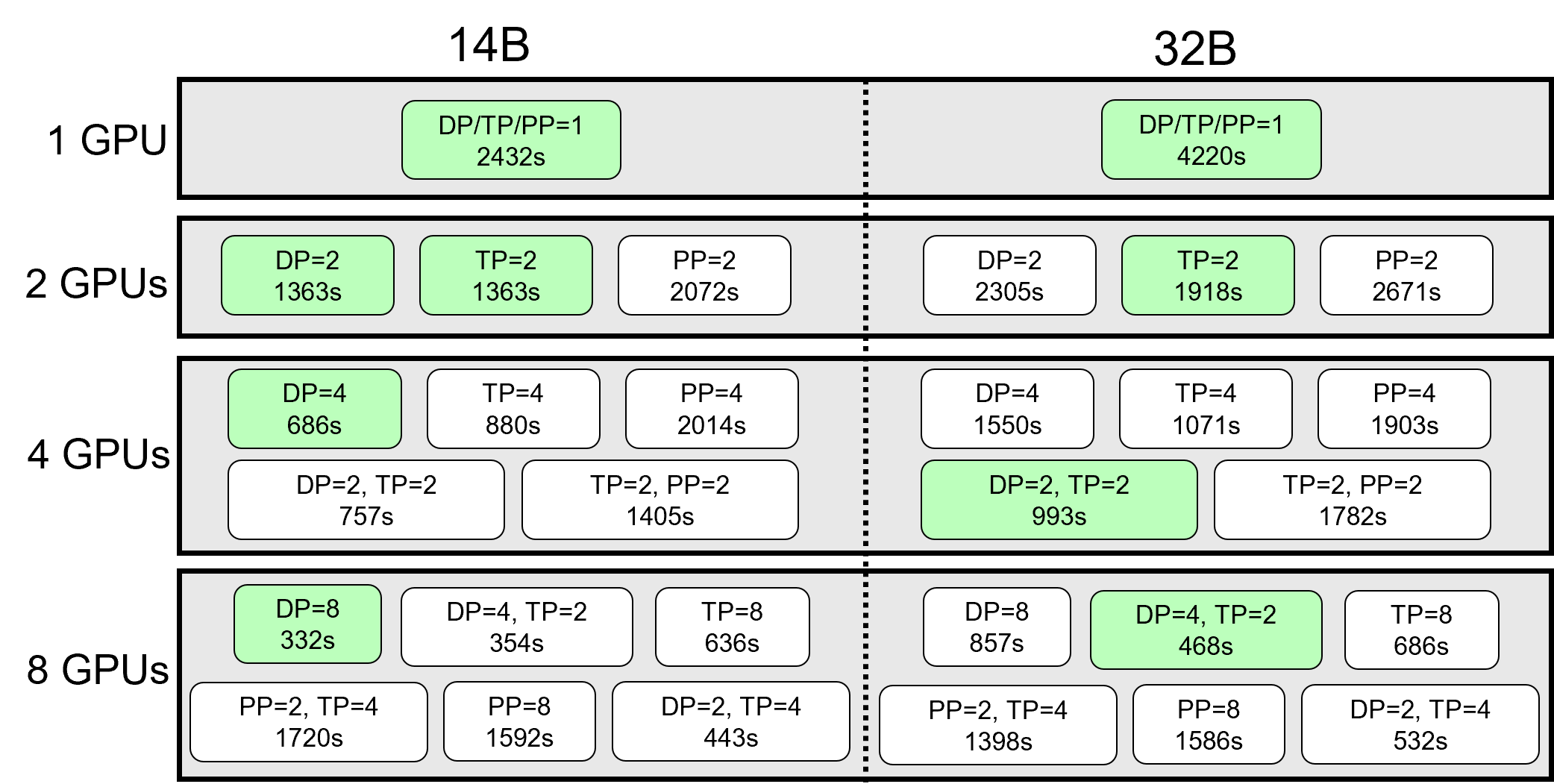}
    \caption{Mixed config scaling for small models (2k BS).}
    \label{fig:figure-1-small-mixed-inf-duration}
    \vspace{-10pt}
\end{figure}

\subsection{The Limits of Data Parallelism (DP)}
Data Parallelism (DP) is often the default scaling strategy as it avoids inter-GPU communication. We investigate its efficacy by scaling the 8B model from 1 to 8 GPUs (Figure~\ref{fig:figure7-8b-scaling}). This setup provides an idealized view of near-linear throughput scaling under independent replicas. However, it also exposes inefficiencies arising from request skew and memory imbalance, which limit achievable gains in practice.

\paragraph*{\bf Throughput Linearity vs. Resource Efficiency}
Figure~\ref{fig:figure7-gen} confirms that DP achieves near-linear scaling in aggregate throughput ($4\times$ gains from 2 to 8 GPUs). However, this masks a critical inefficiency: the system achieves throughput by processing more streams of requests in parallel on different DP replicas, not by accelerating the throughput of any request.
\begin{itemize}[topsep=0pt,itemsep=0pt,leftmargin=12pt] 
    \item \textbf{Stranded Capacity:} DP relies on a shared-nothing memory architecture, where a request on GPU 0 cannot access free memory on GPU 1. This creates ``stranded capacity'', where one replica thrashes (triggering preemption) while another has idle pages. The imbalance worsens under skewed request arrival patterns and dynamic KV growth, leading to underutilized HBM and degraded throughput/latency efficiency across the cluster.
    \item \textbf{Bandwidth Interference:} Figure~\ref{fig:figure7-bw} shows HBM bandwidth oscillating violently (40\%--85\%) rather than staying saturated. This ``sawtooth'' pattern represents the forced interleaving of compute-bound prefills (from the waiting queue) with memory-bound decodes. Because DP replicas are capacity-constrained (Figure~\ref{fig:figure7-kv}), they must constantly ingest new requests to fill voids left by completed ones, causing resource contention that destabilizes tail latency. The capacity contention triggers request preemption (Figure~\ref{fig:figure7-reqs}) with lower degrees of DP experiencing throttling. 
\end{itemize}

\begin{observationbox}
\textbf{Observation 4:} 
DP replicates model weights and partitions requests rather than pooling memory, and does not increase the KV capacity available to an individual long-running request. For reasoning workloads with long and variable chain-of-thought lengths, tail latency is therefore dominated by the replica that reaches KV saturation first. This indicates that DP-only scaling is insufficient when performance is limited by per-request KV growth; memory pooling, TP, or KV offload is required to increase the effective capacity margin.
\end{observationbox}

\section{Analysis II: 3D Parallelism for Large Models}
\label{sec:analysis_large}

While Analysis I (\S~\ref{sec:analysis_small}) characterized the ``Capacity Trap'' inherent to DP, this section investigates the "What-If" scenario: \textit{Do alternative parallelism strategies like TP or PP offer acceleration remedies, or do they introduce new bottlenecks as models scale?} We trace the scaling efficacy from 14B to 671B parameters to uncover divergences in optimal strategies.

\subsection{From DP to TP Scaling}
\label{sec:tp_scaling}
\begin{figure*}
    \centering
    \begin{minipage}{0.32\textwidth}
        \centering
        \includegraphics[width=\linewidth]{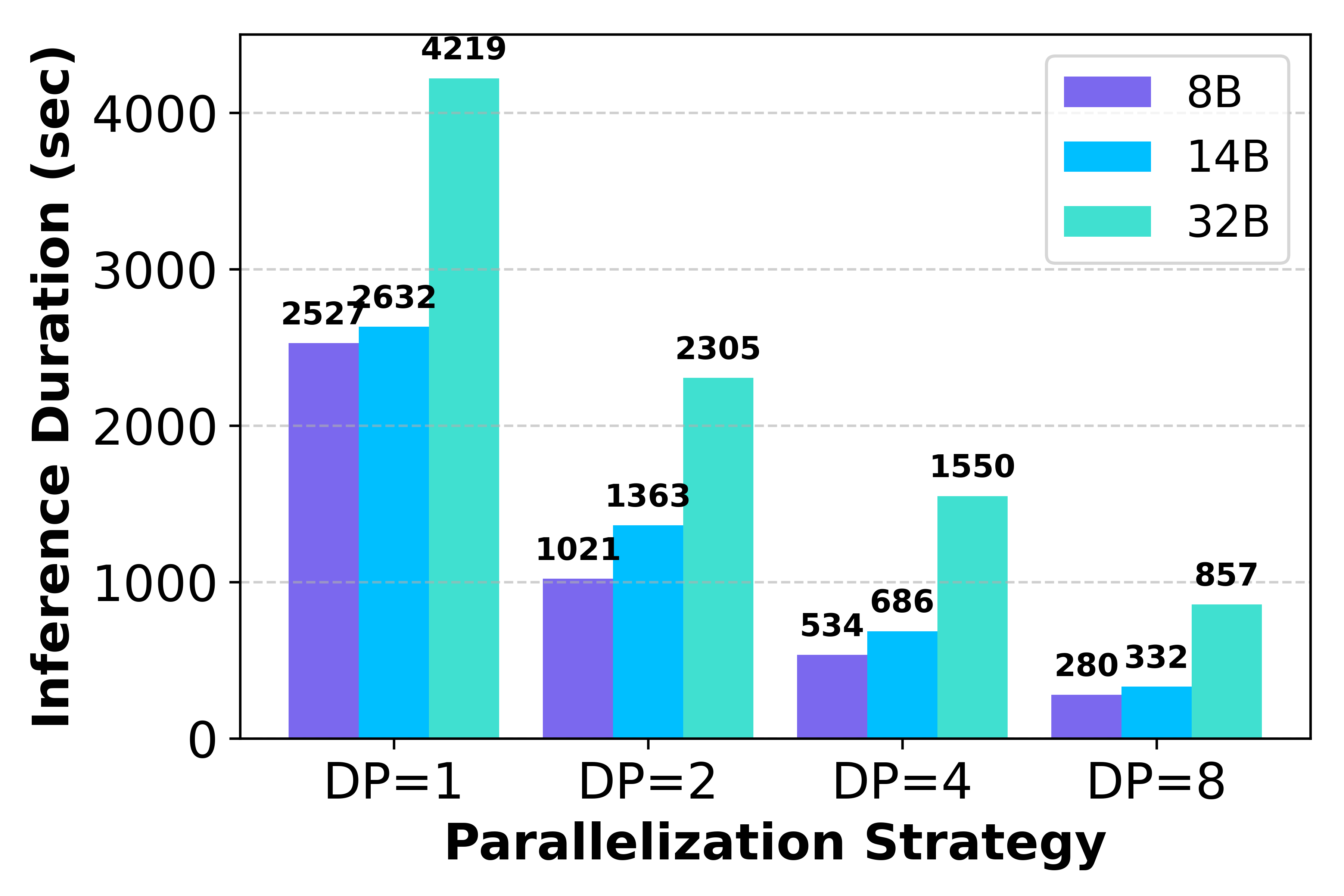}
        \caption{DP Scaling for small models.}
        \label{fig:figure-1-dp-small-inf-duration}
    \end{minipage}
    \hfill
    \begin{minipage}{0.32\textwidth}
        \centering
        \includegraphics[width=\linewidth]{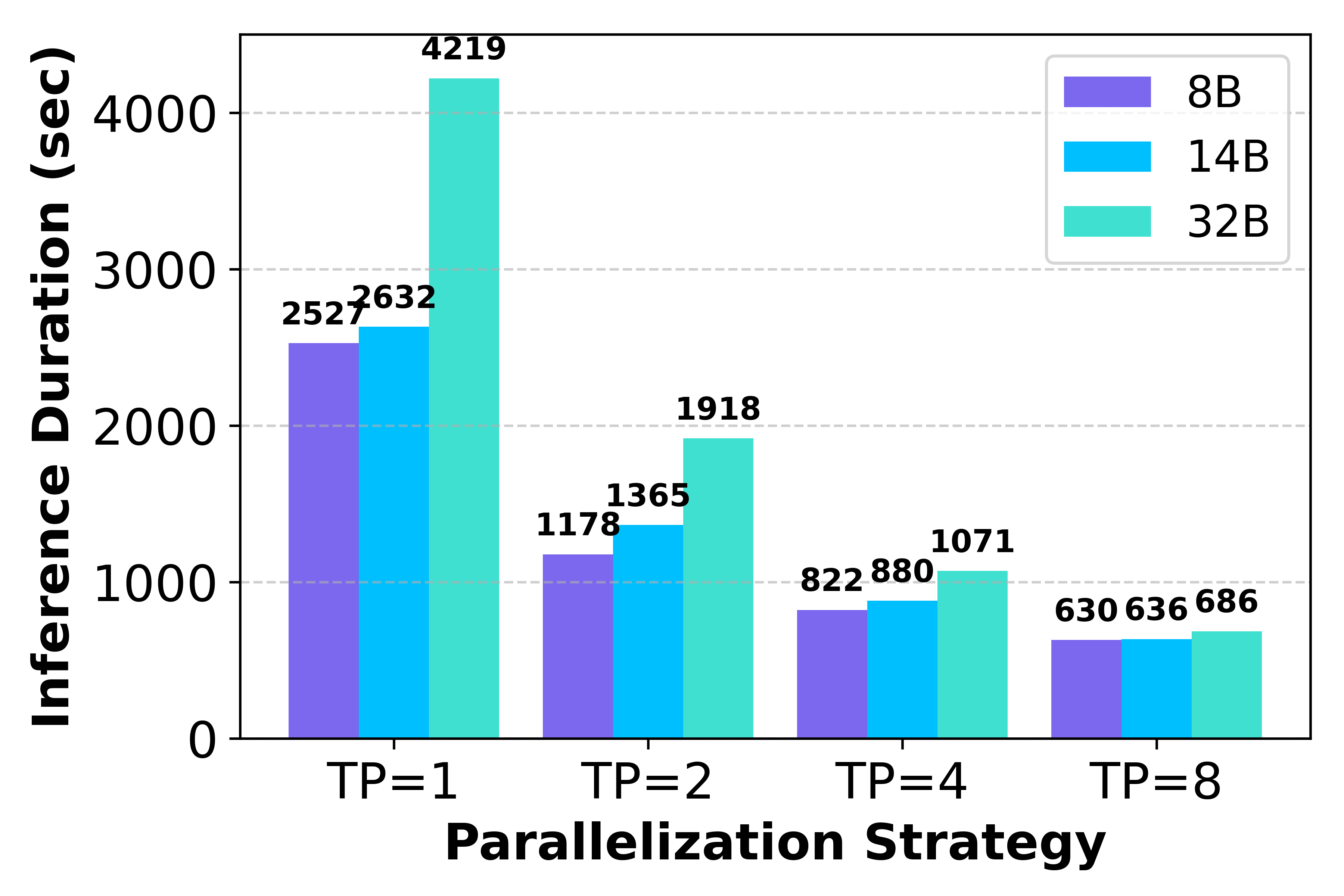}
        \caption{TP Scaling for small models.}
        \label{fig:figure-1-tp-small-inf-duration}
    \end{minipage}
    \hfill
    \begin{minipage}{0.32\textwidth}
        \centering
        \includegraphics[width=\linewidth]{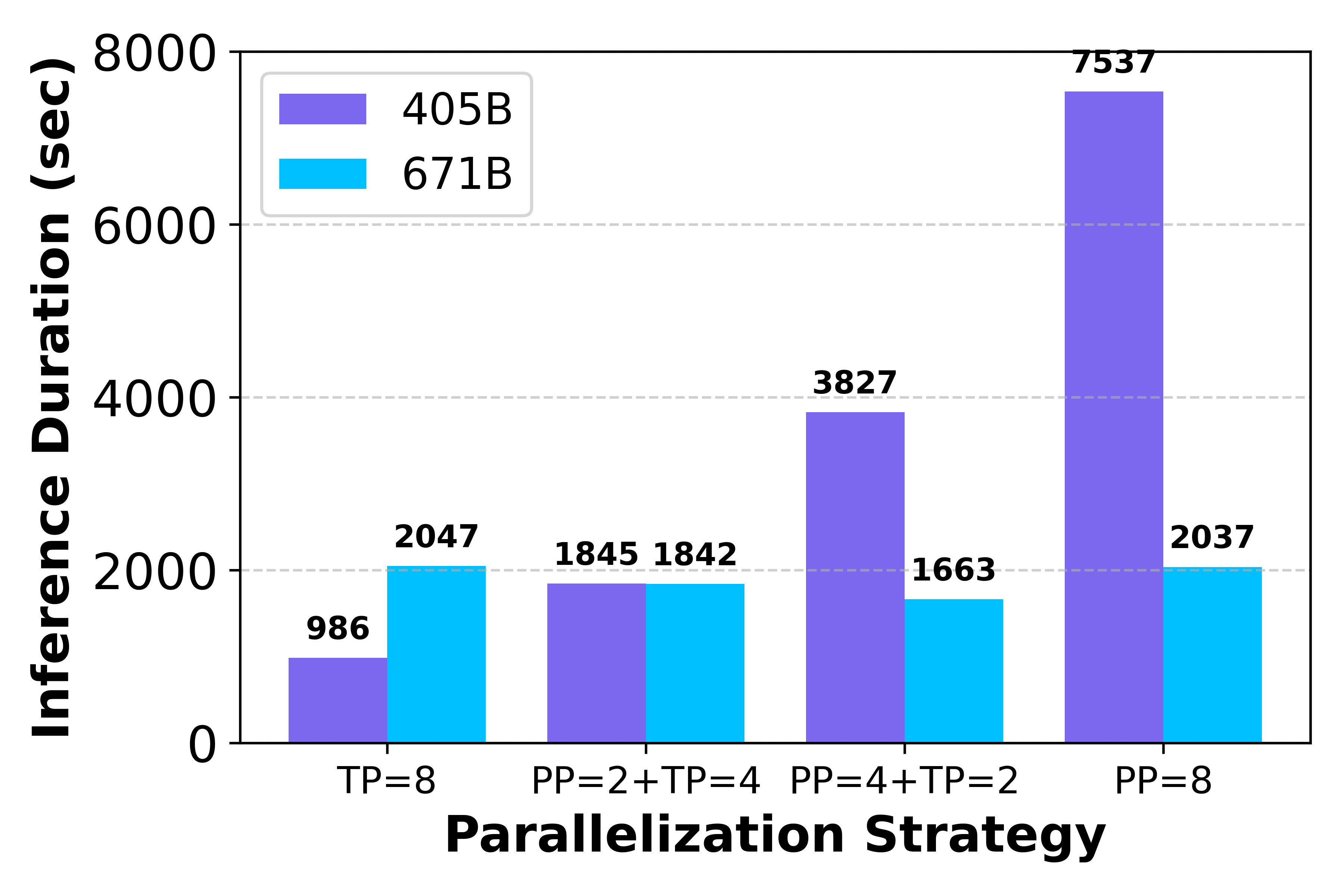}
        \caption{E2E for large models (2k BS).}
        \label{fig:figure-1-large-inf-duration}
    \end{minipage}
\end{figure*}

To determine where DP fails, we compare its scaling efficiency against Tensor Parallelism (TP) for small-to-medium models (Figure~\ref{fig:figure-1-tp-small-inf-duration} vs. Figure~\ref{fig:figure-1-dp-small-inf-duration}) with batch size of 2K requests.

\paragraph*{\bf DP vs TP Inflection Point}
For the 8B and 14B models, TP introduces communication overhead that outweighs its benefits. However, the 32B model reveals a critical architectural inflection point:

\begin{itemize}[topsep=0pt,itemsep=0pt,leftmargin=12pt] 
    \item \textbf{Sublinear DP vs. higher-efficiency TP:} As shown in Figure~\ref{fig:figure-1-dp-small-inf-duration}, DP scaling for the 32B model yields diminishing returns, achieving a $4.9\times$ speedup on 8 GPUs. In contrast, Figure~\ref{fig:figure-1-tp-small-inf-duration} shows that TP achieves a higher, though still sublinear, speedup of \textbf{$6.15\times$} by reducing inference duration from 4219s to 686s.
    \item \textbf{Why TP Wins (The Capacity Release):} The 32B model weights consume $\approx$64 GB in FP16. In a DP configuration, every GPU must replicate this 64 GB, leaving only $\approx$77 GB of the 141 GB HBM for KV cache. This creates a hard capacity ceiling that forces preemption. In contrast, TP shards the weights; at TP=8, the model consumes only $\approx$8 GB per GPU, freeing up $\approx$133 GB per device for KV cache. This massive release of HBM capacity allows TP to sustain a much larger active batch size without preemption, eliminating the re-computation overhead that throttles DP.
\end{itemize}

\begin{figure*}[!t]
    \centering
    \begin{subfigure}[h!]{0.24\textwidth}
        \centering
        \includegraphics[width=\linewidth]{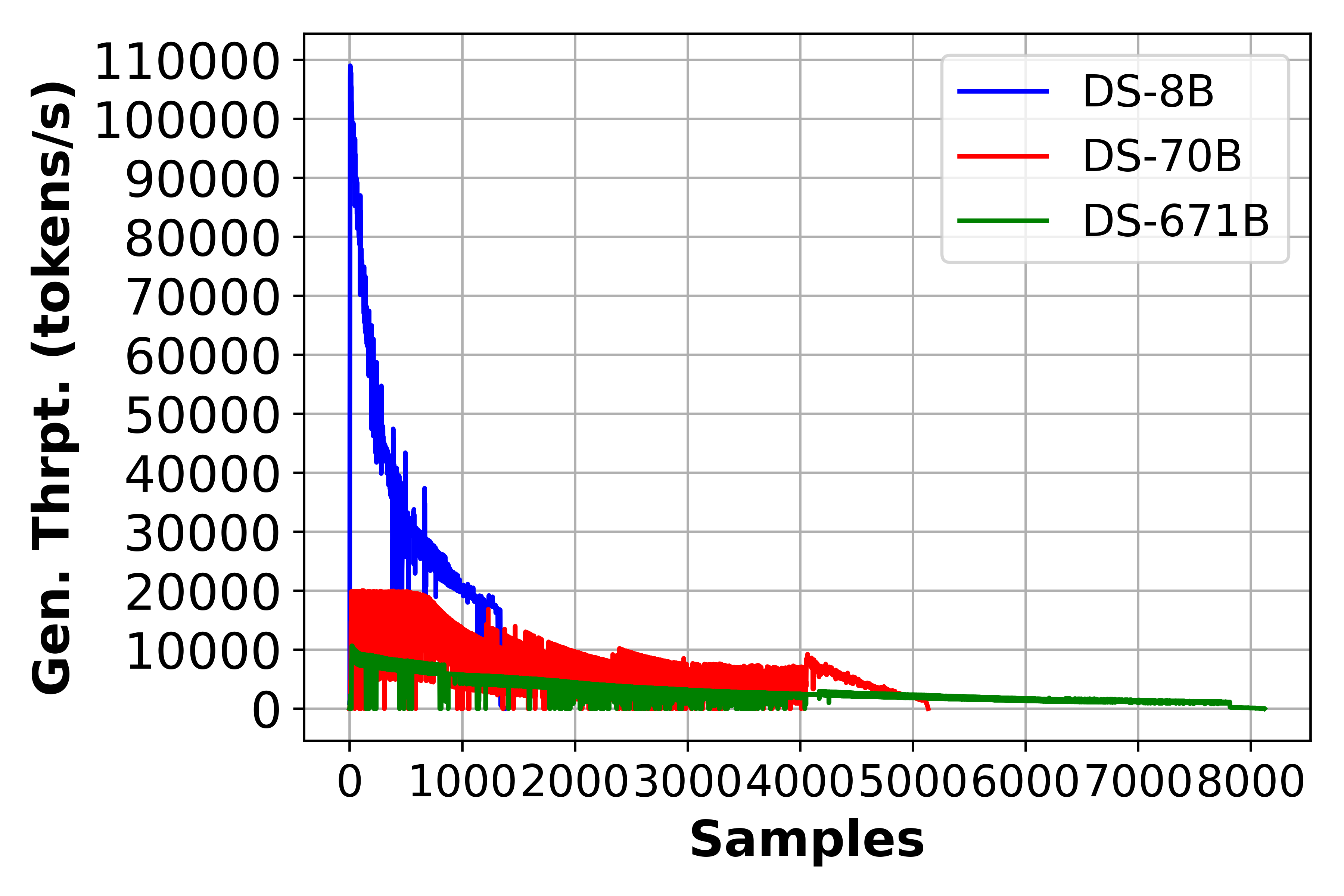}
        \caption{Generation Throughput}
        \label{fig:figure3-gen}
    \end{subfigure}
    \hfill
    \begin{subfigure}[h!]{0.24\textwidth}
        \centering
        \includegraphics[width=\linewidth]{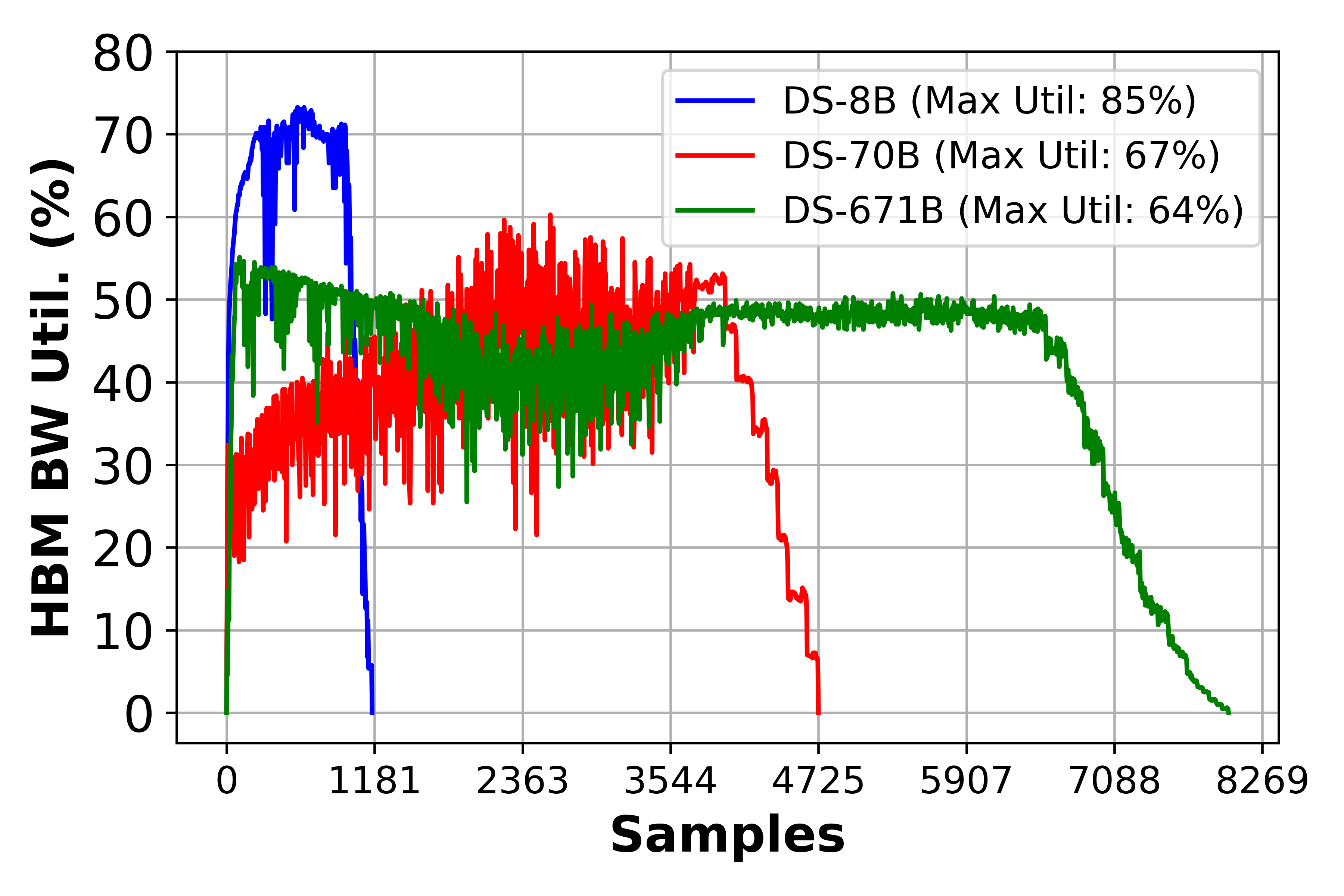}
        \caption{Average HBM BW Util.}
        \label{fig:figure3-bw}
    \end{subfigure}
    \hfill
    \begin{subfigure}[h!]{0.24\textwidth}
        \centering
        \includegraphics[width=\linewidth]{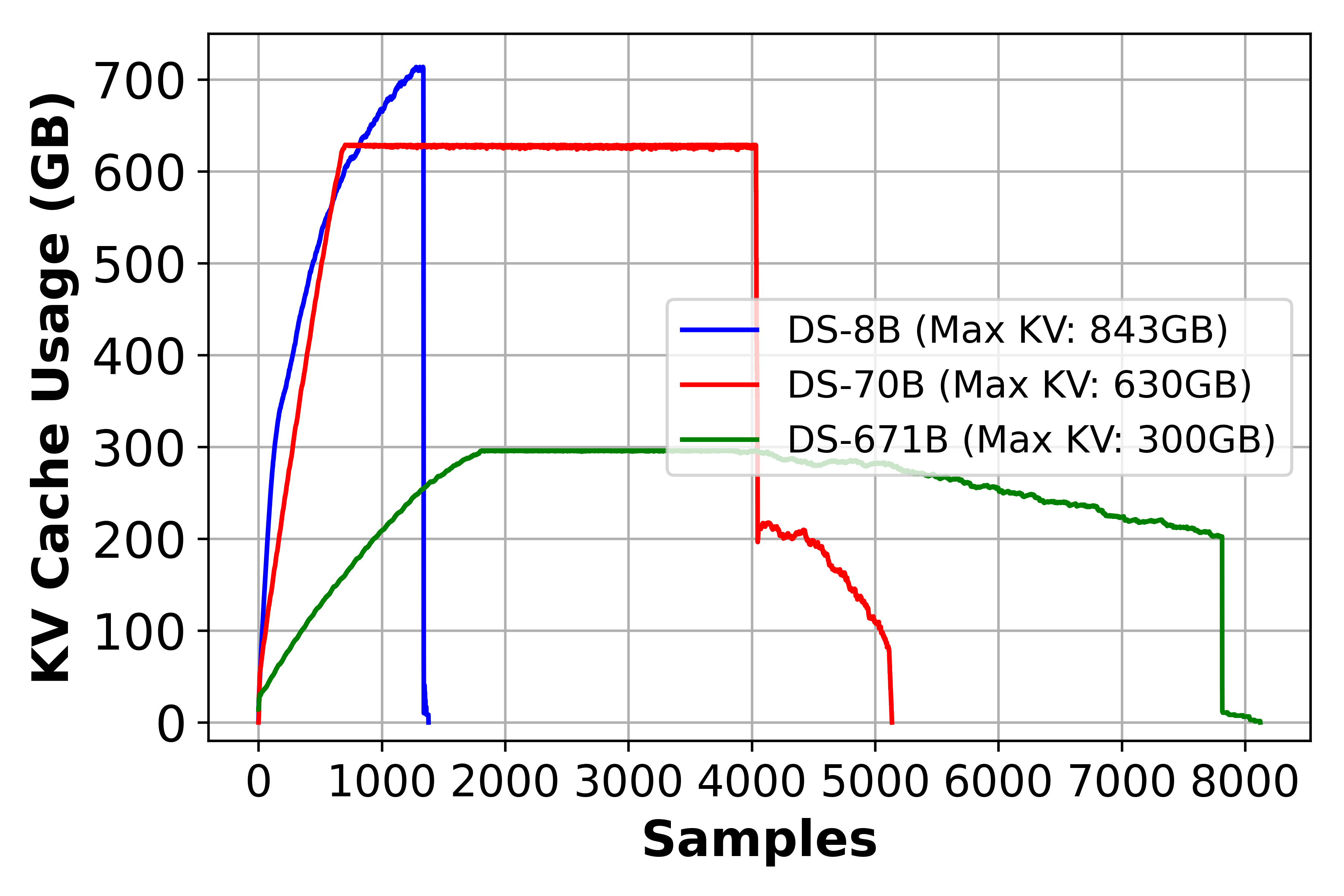}
        \caption{Aggregated KV Cache Util.}
        \label{fig:figure3-kv}
    \end{subfigure}
    \hfill
    \begin{subfigure}[h!]{0.24\textwidth}
        \centering
        \includegraphics[width=\linewidth]{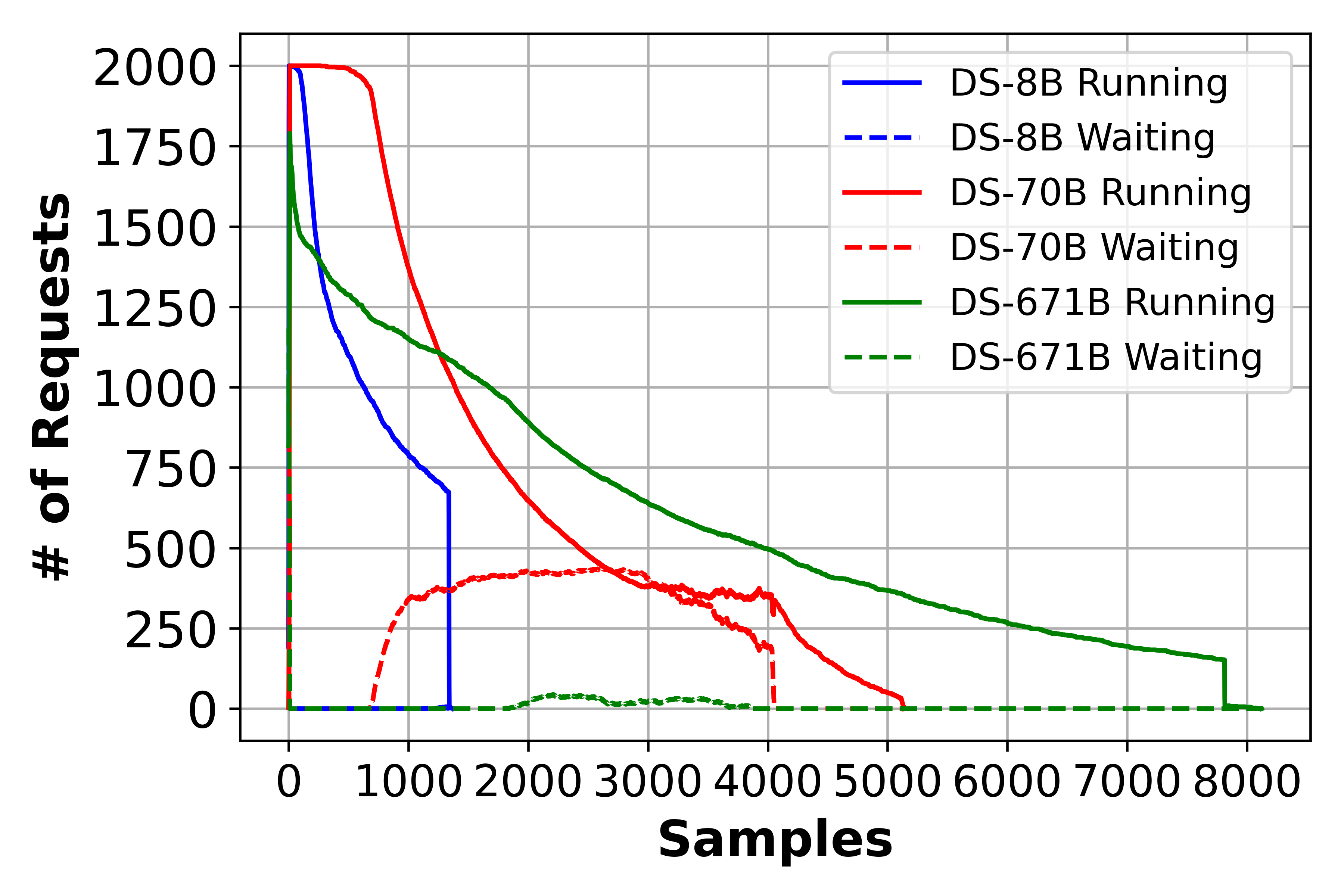}
        \caption{Request Analysis}
        \label{fig:figure3-reqs}
    \end{subfigure}
    \caption{Model parameter scaling on 8x H200 GPUs.}
    \label{fig:figure3-model-scaling}
\end{figure*}

These results indicate that the benefit of TP arises primarily from released KV capacity rather than improved kernel efficiency, as the performance inflection aligns with the point at which DP replicas become KV-capacity-bound. This suggests that memory headroom, not compute efficiency, is the dominant scaling bottleneck, and that TP’s gains stem from alleviating KV pressure rather than accelerating execution.

\subsection{Hybrid Parallelism Sweet Spot for Small Models}

\label{sec:hybrid_small}
Figure~\ref{fig:figure-1-small-mixed-inf-duration} explores the trade-off space of hybrid configurations (DP+TP, TP+PP) for 14B and 32B models on a fixed 8-GPU budget. It highlights how combining parallelism strategies balances communication overhead and HBM efficiency, revealing regimes where hybrid schemes outperform pure DP or TP by reducing preemption while maintaining scalable throughput.

\paragraph*{\bf Balancing Communication Overhead vs. Memory Capacity}
The data reveals that optimal performance requires balancing per-device latency (TP) with cluster-level concurrency (DP).
\begin{itemize}[topsep=0pt,itemsep=0pt,leftmargin=12pt] 
    \item \textbf{14B (DP Dominant):} The optimal configuration is pure DP=8 (332s). Hybrid strategies like $PP=2+TP=4$ (1172s) are $\approx 3.5\times$ slower. The model is too small to justify the synchronization costs of model parallelism; simply running more independent streams is most efficient.
    \item \textbf{32B (Hybrid Dominant):} The optimal configuration is neither pure DP (857s) nor pure TP (686s), but the hybrid \textbf{DP=4 + TP=2} (484s). This configuration uses minimal TP ($TP=2$) to alleviate the per-device capacity/bandwidth bottleneck to prevent preemption, while dedicating the remaining scaling factor ($DP=4$) to concurrency. This ``Right-Sized'' TP approach outperforms pure TP by $\approx~30\%$, proving that for medium models, minimizing communication degree is as critical as maximizing bandwidth.
\end{itemize}

\begin{observationbox}
\textbf{Observation 5:} 
The preferred parallelism strategy depends on whether the dominant overhead is communication or capacity-induced preemption. Smaller models, such as 14B, favor DP because their weights fit comfortably within per-GPU HBM and TP communication overhead is not justified. For larger models, such as 32B, TP becomes beneficial because sharding model weights releases HBM for KV cache. Once the cost of DP preemption and recomputation exceeds TP's NVLink communication overhead, TP becomes the better scaling strategy.
\end{observationbox}

\begin{figure*}[!t]
    \centering
    \begin{subfigure}[b]{0.3\textwidth}
        \centering
        \includegraphics[width=\linewidth]{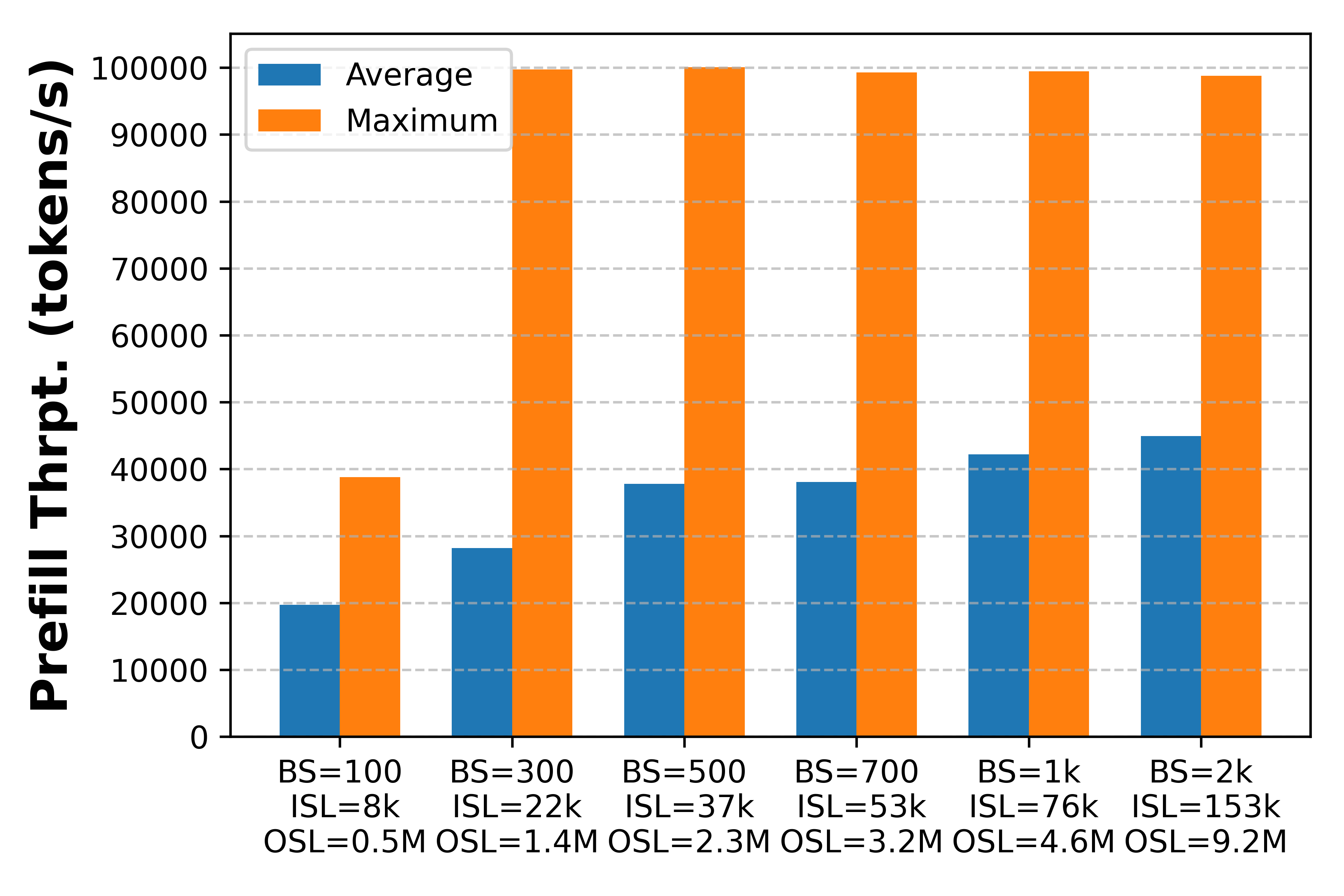}
        \caption{Prefill Phase Throughput}
        \label{fig:figure4-prefill-thrpt}
    \end{subfigure}\hspace{0.01\textwidth}
    \begin{subfigure}[b]{0.3\textwidth}
        \centering
        \includegraphics[width=\linewidth]{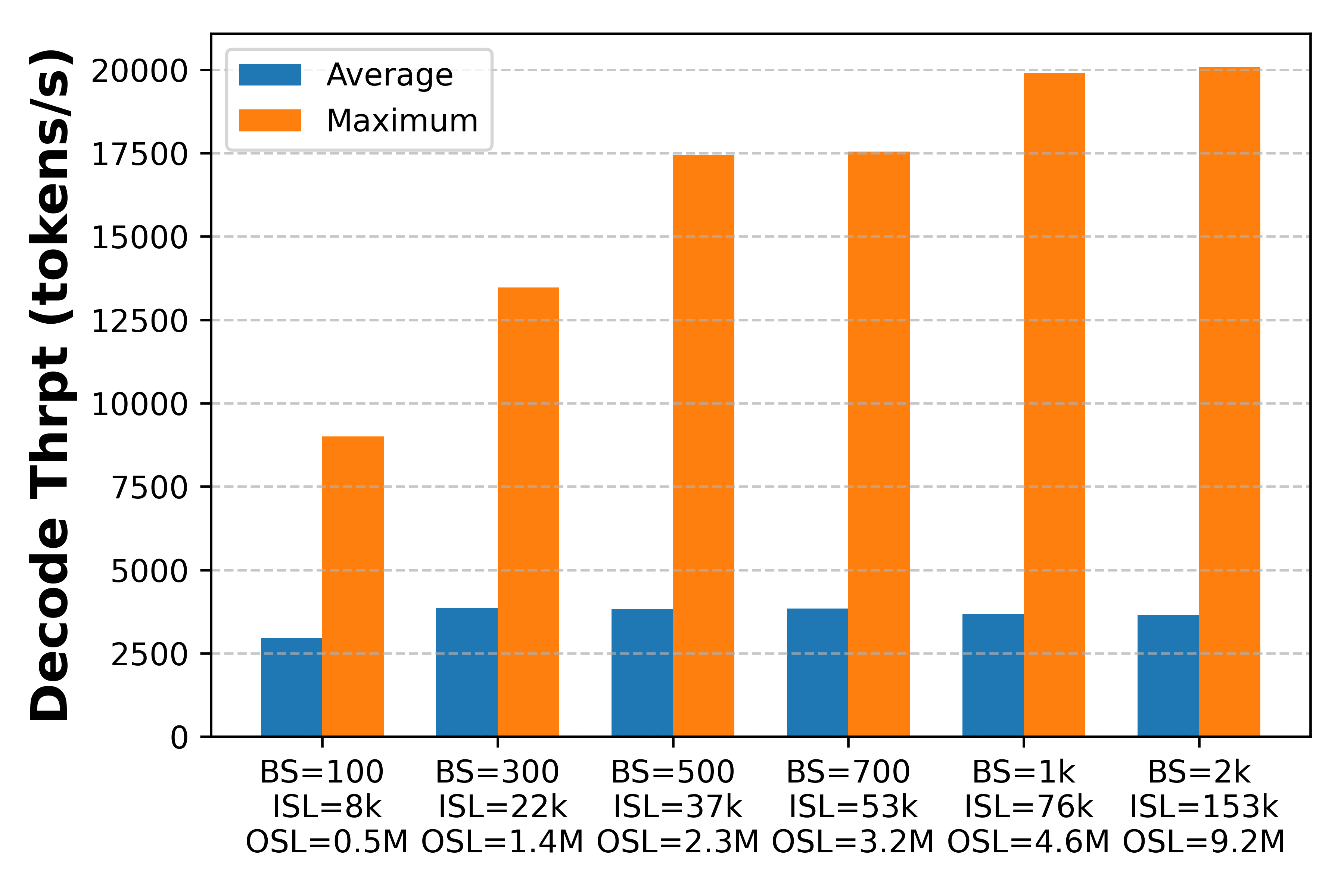}
        \caption{Decode Phase Throughput}
        \label{fig:figure4-decode-thrpt}
    \end{subfigure}\hspace{0.01\textwidth}
    \begin{subfigure}[b]{0.3\textwidth}
        \centering
        \includegraphics[width=\linewidth]{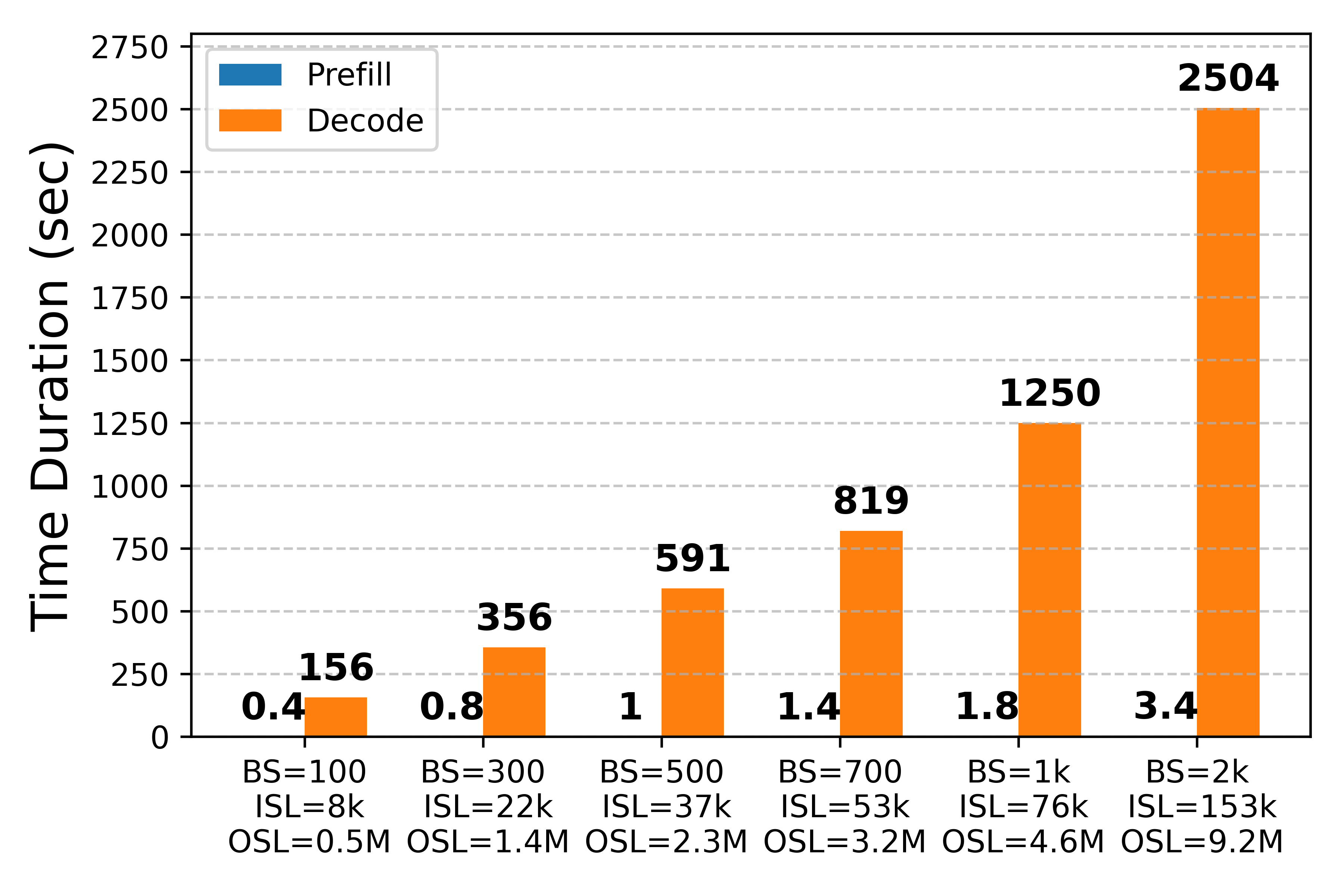}
        \caption{Prefill and Decode Duration}
        \label{fig:figure4-duration}
    \end{subfigure}\hspace{0.01\textwidth}
    \begin{subfigure}[b]{0.48\textwidth}
        \centering
        \includegraphics[width=\linewidth]{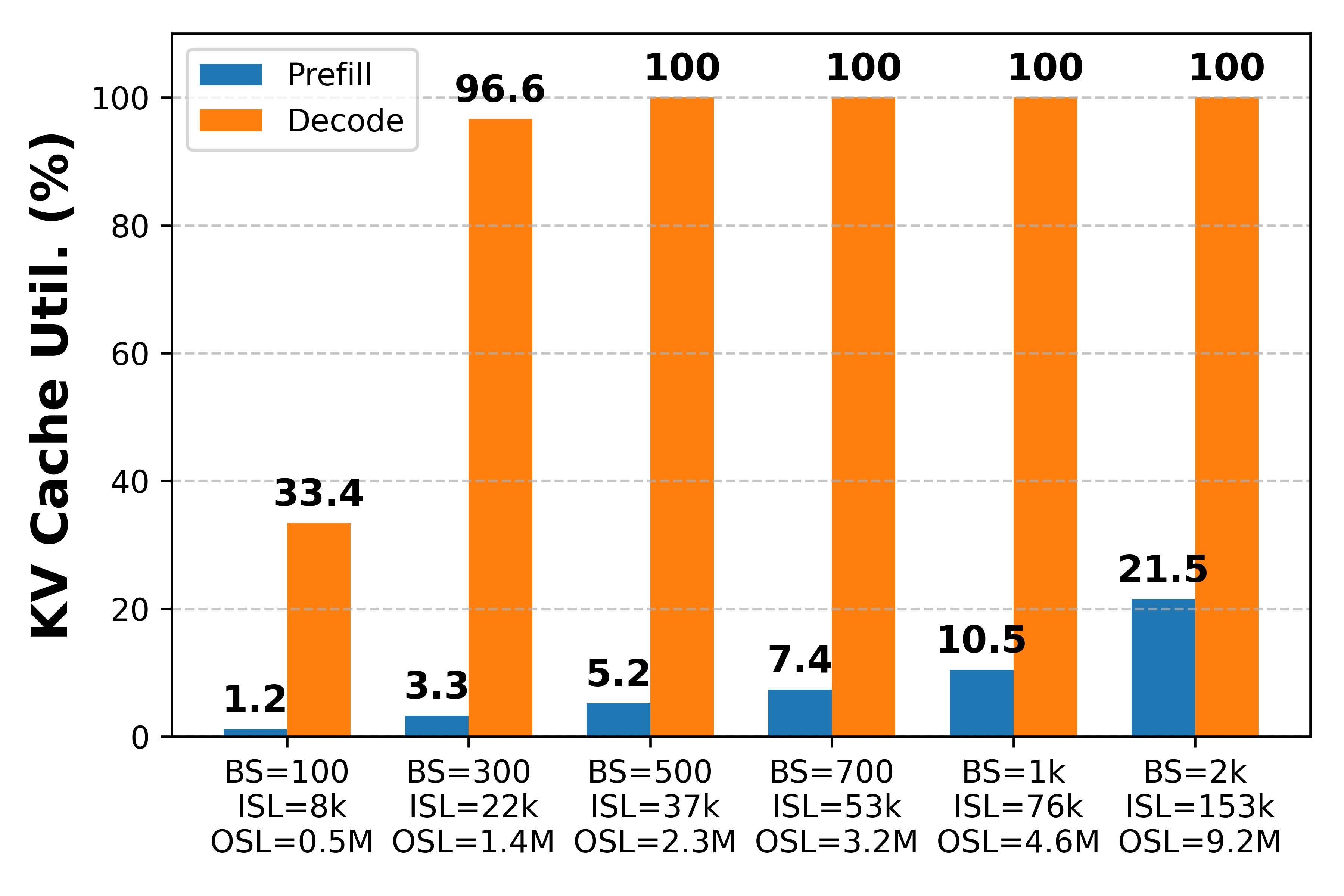}
        \caption{Prefill and Decode KV Util.}
        \label{fig:figure4-kv}
    \end{subfigure}\hspace{0.005\textwidth}
    \begin{subfigure}[b]{0.48\textwidth}
        \centering
        \includegraphics[width=\linewidth]{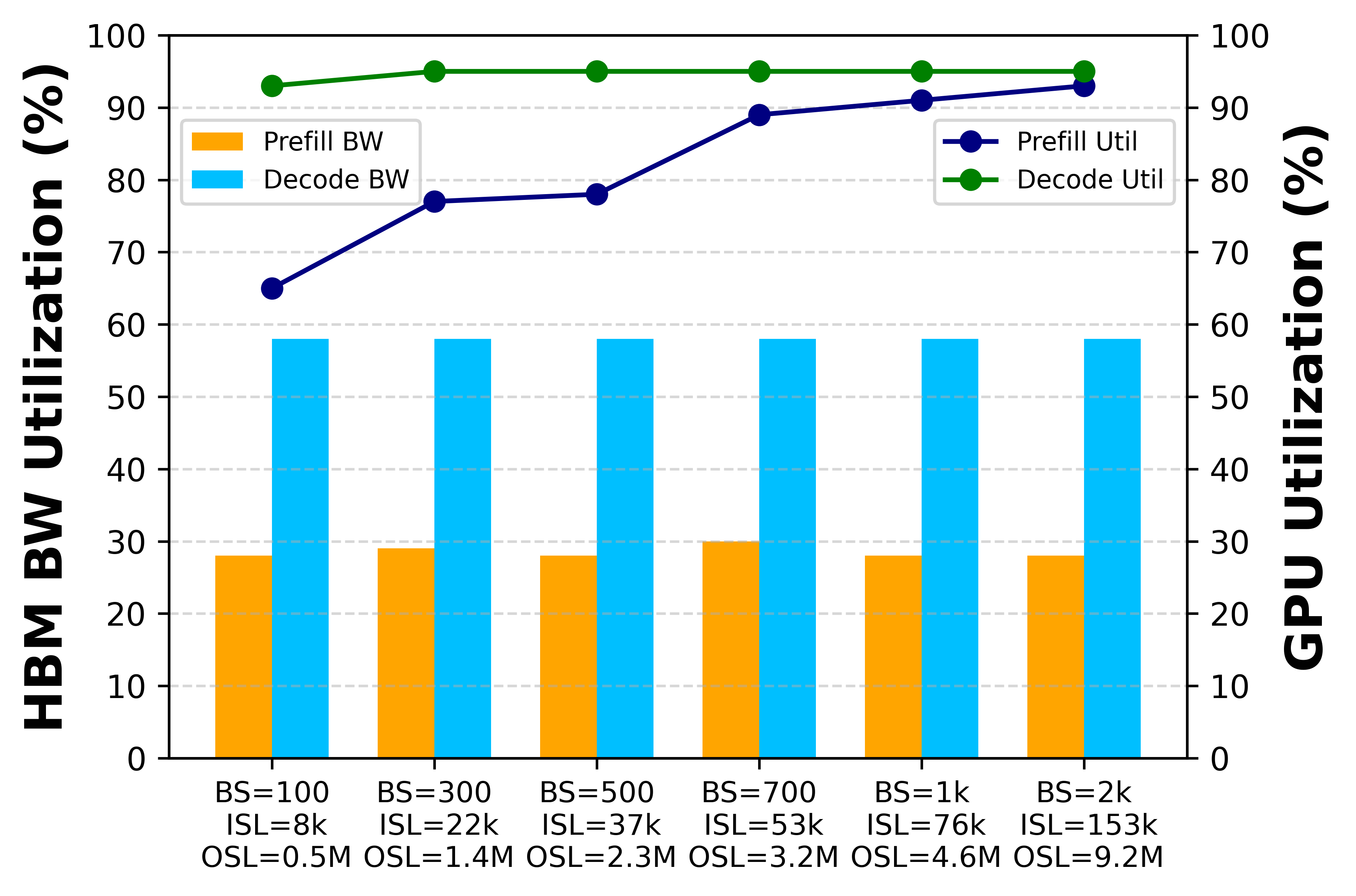}
        \caption{Prefill and Decode GPU Util.}
        \label{fig:figure4-gpu-hbm-util}
    \end{subfigure}

    \caption{Analysis of Prefill and Decode Phase during AI Inference.}
    \label{fig:figure4-prefill-decode}
\end{figure*}

\subsection{Frontier Scaling: Dense vs. MoE Divergence}
\label{sec:frontier_scaling}

Scaling to frontier-class models (405B and 671B) requires adaptation of ``Parallelism'' based on model architecture. As shown in Figure~\ref{fig:figure-1-large-inf-duration}, there is no single universal strategy; instead, the optimal approach varies based on model density.

\paragraph*{\bf Llama-405B (Dense)}
For the dense Llama-405B, due to large KV footprint the combined memory capacity of all GPUs is needed limiting DP=1. TP=8 or PP=8 are the only viable strategies, with TP=8 completing in 986s, while PP=8 is catastrophic (7537s).
\begin{itemize}[topsep=0pt,itemsep=0pt,leftmargin=12pt] 
    \item \textbf{The Penalty of Density:} Llama-405B activates all parameters for every token. This creates a massive computational workload that should theoretically hide pipeline bubbles. However, the $7.6\times$ slowdown of PP suggests the opposite. The massive KV state of 405B (1.05 MB/token) restricts the number of micro-batches that can fit in memory, forcing the pipeline to run with large bubbles (idle time). Despite high arithmetic intensity, dense models such as Llama-405B cannot amortize pipeline bubbles due to insufficient KV capacity per stage, causing PP to exacerbate rather than alleviate decode-phase stalls.
    \item \textbf{Inter-Stage Traffic:} Furthermore, moving the dense activation tensors between PP stages saturates the interconnect. The TP=8 configuration avoids this by keeping activations node-local and only exchanging reduced gradients/activations over NVLink, which offers higher bandwidth than the PCIe/Link paths often used for PP stage transitions.
\end{itemize}

\paragraph*{\bf DeepSeek-R1-671B (MoE)}
In stark contrast, the sparse DeepSeek-R1 model favors a Hybrid PP=4 + TP=2 strategy (1663s), beating pure TP=8 (2047s).
\begin{itemize}[topsep=0pt,itemsep=0pt,leftmargin=12pt]
    \item \textbf{Sensitivity to Synchronization:} While R1 is larger (671B), it is a MoE model with only $\approx$37B active parameters. This low active parameter count means the compute-to-communication ratio is much lower than 405B. In a TP=8 setup, the constant all-reduce synchronization becomes a bottleneck because the GPU spends less time computing between syncs.
    \item \textbf{The MLA Advantage:} R1 employs MLA, which significantly compresses KV cache. This architectural feature works synergistically with PP. The reduced memory footprint allows R1 to support a higher micro-batch depth than 405B, effectively filling the pipeline bubbles.
    \item \textbf{Optimal Balance:} With $PP=4$, R1 splits the model into smaller stages that fit in memory, and by using $TP=2$, it minimizes pipeline bubble overhead while using combined GPU capacity. This result ($1663s$ vs $2047s$) confirms that for sparse reasoning models, minimizing TP degree is preferred over aggregating capacity and bandwidth.
\end{itemize}

\begin{observationbox}
\textbf{Observation 6:} 
Dense models benefit from higher-degree TP when inference is limited by HBM bandwidth and model-state footprint, since TP aggregates both memory capacity and bandwidth across GPUs. In contrast, sparse MoE models can become more sensitive to synchronization and pipeline imbalance; therefore, hybrid strategies with higher PP and lower TP may be preferable when TP communication dominates. This suggests that serving systems should choose DP/TP/PP configurations using model-specific profiling rather than a fixed parallelism policy.
\end{observationbox}

\subsection{Impact of Scaling Model Parameters}
To isolate the impact of model scale on inference dynamics, we evaluate three distinct model classes---DeepSeek-8B (Small/Dense), DeepSeek-70B (Medium/Dense), and DeepSeek-R1-671B (Frontier/Sparse)---on a fixed hardware budget of 8$\times$H200 GPUs. Each model uses its optimal parallelization strategy: pure DP for 8B, Hybrid TP for 70B, and Hybrid PP+TP for 671B.

\paragraph*{\bf The Sublinear Throughput Degradation}
Figure~\ref{fig:figure3-gen} illustrates the generation throughput. As expected, the peak throughput drops as parameter count increases, but the degradation is notably \textit{sublinear}. A $9\times$ parameter increase from 8B to 70B results in only a $5\times$--$6\times$ drop in peak throughput. This efficiency gain is driven by Tensor Parallelism (used for 70B), which aggregates memory bandwidth across GPUs, partially offsetting the increased FLOPs requirement. However, the 671B model (green curve) exhibits a long, flat tail, reflecting the extended "reasoning" nature of its output compared to the bursty completion of the smaller distilled models.

\paragraph*{\bf The Bandwidth-Compute Inversion}
Telemetry reveals a fundamental shift in bottlenecks as models scale:
\begin{itemize}[topsep=0pt,itemsep=0pt,leftmargin=12pt]
    \item \textbf{Small Models (8B):} Exhibit the highest HBM utilization ($\approx$85\%, Figure~\ref{fig:figure3-bw}). The workload is memory-bandwidth bound because the small weight matrices are loaded rapidly, saturating the bus.
    \item \textbf{Frontier Models (671B):} Despite their size, they show \textit{lower} average HBM utilization ($\approx$50--60\%). This counter-intuitive result confirms that ultra-large sparse models are bound by \textbf{synchronization and routing latency}, not raw memory bandwidth. The overhead of Pipeline Parallelism bubbles and MoE expert routing prevents the system from saturating the HBM links.
\end{itemize}

\paragraph*{\bf The MLA Anomaly in Capacity}
Figure~\ref{fig:figure3-kv} highlights a critical architectural advantage of DeepSeek-R1. Despite having $10\times$ the parameters of the 70B model, its rate of KV-cache consumption is surprisingly moderate. This is due to Multi-Head Latent Attention (MLA), which compresses the KV state. In contrast, the dense 70B model (red line) aggressively consumes capacity, reaching its ceiling faster relative to its throughput resulting in request throttling (Figure~\ref{fig:figure3-reqs}). This suggests that for future reasoning systems, architectural compression (like MLA) is as critical as hardware capacity for sustaining long-context inference.

\section{Analysis III: Prefill vs Decode Resource Requirement Divergence}
\label{sec:analysis_divergence}

The preceding analyses established that reasoning workloads are capacity-constrained (Analysis I) and sensitive to parallelism overheads (Analysis II). This section investigates the underlying physical cause: the extreme \textit{resource divergence} between the Prefill and Decode phases. We characterize the "What-If" scenario: \textit{Can a monolithic accelerator architecture efficiently serve a workload that oscillates between two orthogonal hardware bottlenecks?}

\begin{figure*}[!t]
    \centering
    \begin{subfigure}[b]{0.49\textwidth}
        \centering
        \includegraphics[width=\linewidth]{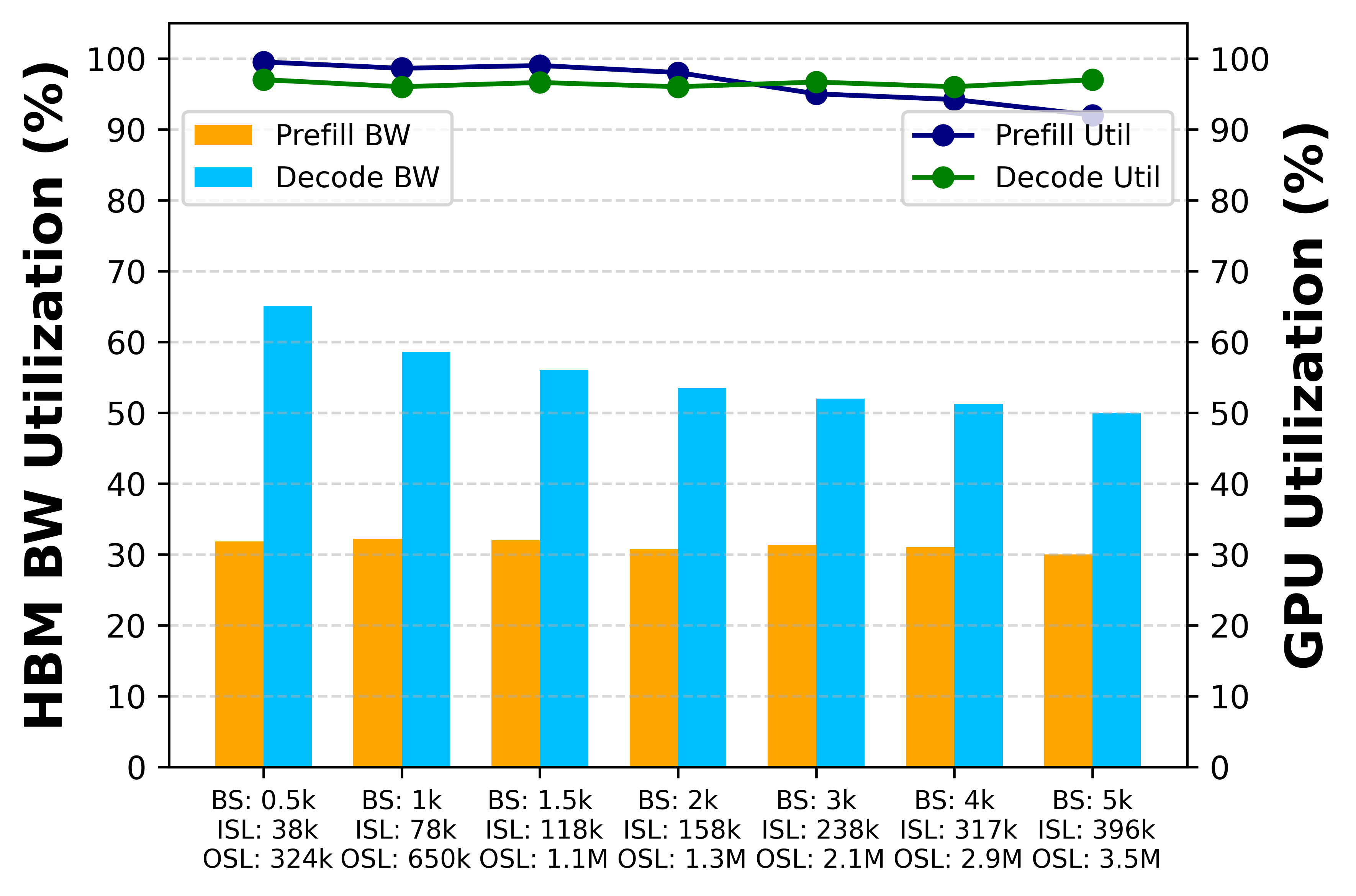}
        \caption{GPU/HBM utilization for Llama405B.}
        \label{fig:figure6-llama}
    \end{subfigure}
    \begin{subfigure}[b]{0.49\textwidth}
        \centering
        \includegraphics[width=\linewidth]{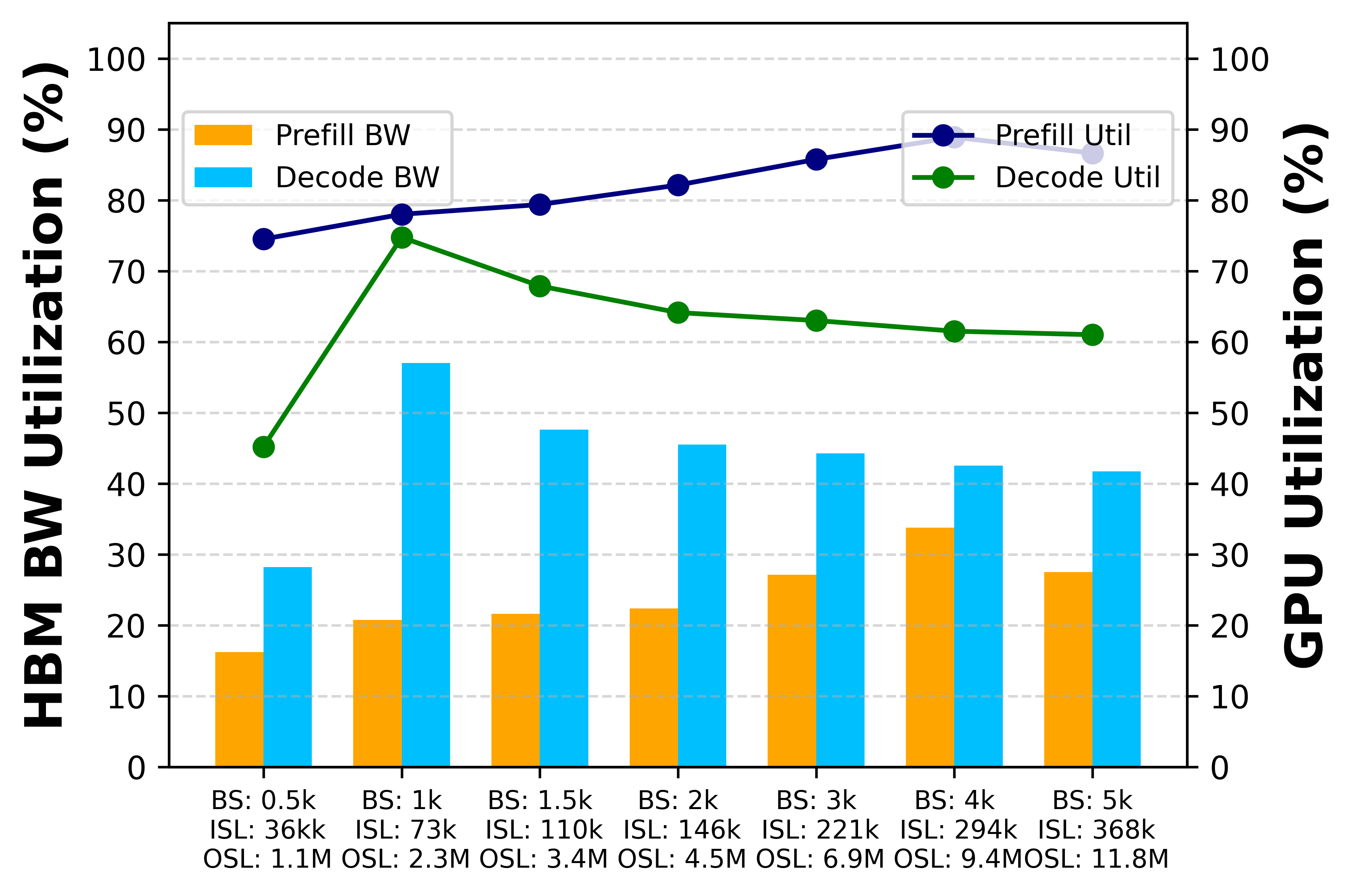}
        \caption{GPU/HBM utilization for DeepSeek-R1-671B.}
        \label{fig:figure6-deepseek}
    \end{subfigure}
    \caption{Analysis of Prefill and Decode resource utilization for varying context lengths.}
    \label{fig:figure6-405B-671B}
\end{figure*}

\begin{figure*}
    \centering
    \begin{subfigure}[b]{0.49\textwidth}
        \centering
        \includegraphics[width=\linewidth]{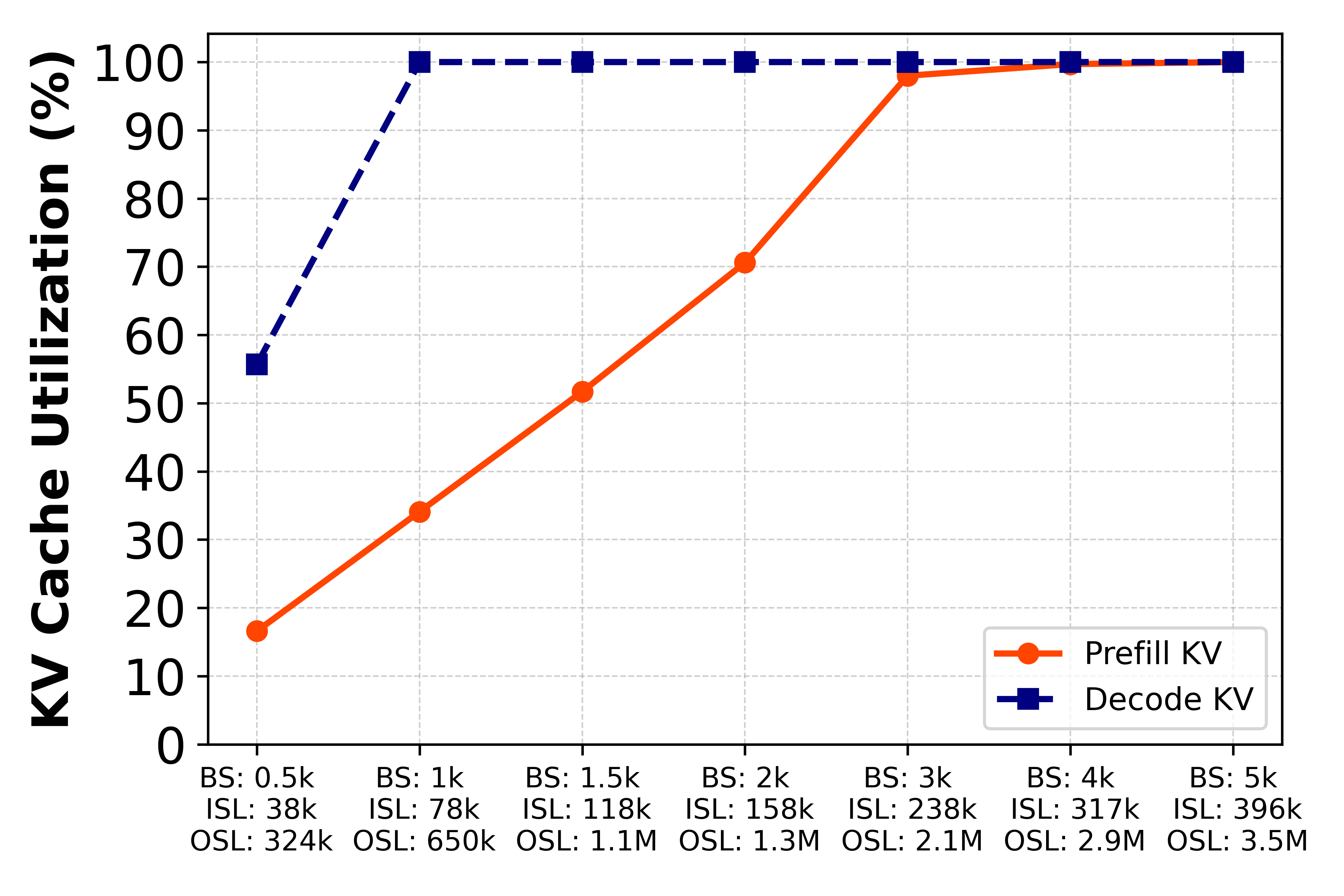}
        \caption{KV Cache utilization.}
        \label{fig:figure6-llama-prefill-decode}
    \end{subfigure}
    \begin{subfigure}[b]{0.49\textwidth}
        \centering
        \includegraphics[width=\linewidth]{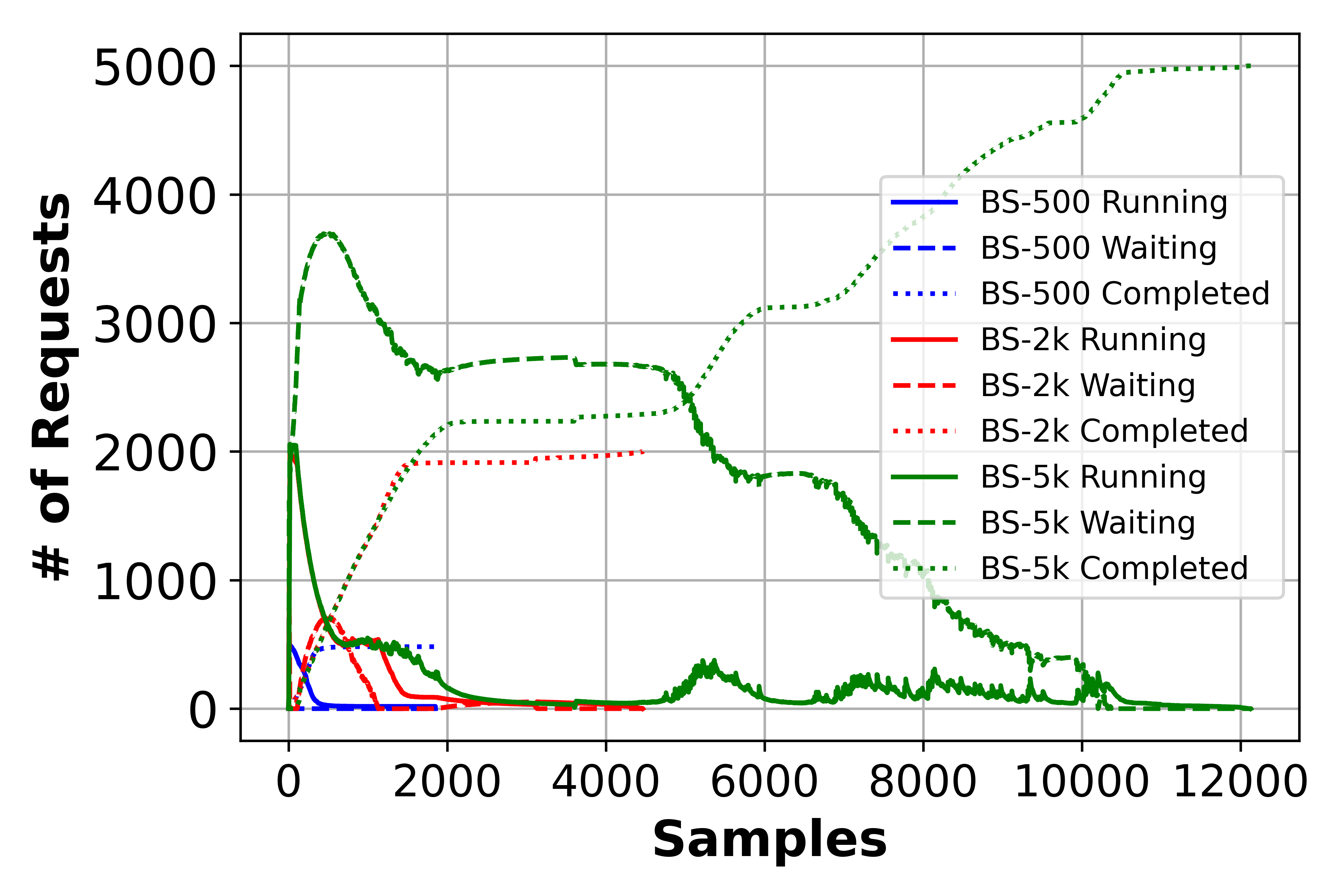}
        \caption{Request Analysis.}
        \label{fig:figure6-llama-req-analysis}
    \end{subfigure}
    \caption{Analysis of Prefill and Decode resource utilization of Llama405B for varying context lengths.}
    \label{fig:figure6-405B}
\end{figure*}

\subsection{Compute vs Memory Bound Phases}
Inference is often treated as a single workload, but telemetry from the 8B model (Figure~\ref{fig:figure4-prefill-decode}) and frontier models (Figure~\ref{fig:figure6-405B-671B}) reveals that the system effectively behaves as two distinct machines. In this series of experiments, we vary the batch size from 100 to 2000 requests to observe how the resource profile evolves. Figures~\ref{fig:figure4-prefill-decode}--\ref{fig:figure5-prefill-decode-kv-proj} collectively trace how resource utilization shifts across inference phases and model scales, revealing that reasoning workloads operate almost entirely in a decode-dominated regime where memory bandwidth and KV capacity, rather than compute, determine the inference throughput and latency.

\paragraph*{\bf The Compute-Bound Prefill}
During the prefill phase, the engine processes all input tokens in parallel. Figure~\ref{fig:figure4-prefill-thrpt} shows high prefill throughput for varying batch sizes showcasing the increase in average throughput with increasing context, while Figure~\ref{fig:figure4-gpu-hbm-util} reveals the corresponding resource signature: SM Occupancy is high, but HBM bandwidth utilization remains low ($\approx$30\%) for the 8B model and $\approx$20\% for the larger 405B and 671B models (Figure~\ref{fig:figure6-deepseek}). This indicates that prefill is \textit{Compute-Bound}. The arithmetic intensity (FLOPs/Byte) is high because the matrix-matrix multiplications (GEMMs) can reuse loaded weights across all tokens in the prompt. In this phase, the H200's 4.8 TB/s bandwidth is underutilized and the KV footprint (Figure~\ref{fig:figure4-kv}) also staying lower and increasing moderately for higher context, effectively leaving performance on the table. Lower prefill utilization (Figure~\ref{fig:figure4-gpu-hbm-util}) for 8B model at small batches arises because prefill is memory/synchronization-limited rather than compute-bound. Frequent kernel launches and cross-GPU synchronization reduce sustained SM occupancy, while short prefill duration (Figure~\ref{fig:figure4-duration}), minimal KV usage (Figure~\ref{fig:figure4-kv}), and moderate HBM bandwidth (Figure~\ref{fig:figure4-gpu-hbm-util}) indicate stall-dominated execution. Decode operates in a steady-state with fewer synchronizations/better locality, leading to higher utilization. As ISL increases, prefill utilization improves due to higher arithmetic intensity from MLP/attention computations.

\paragraph*{\bf The Bandwidth-Bound Decode}
The transition to the decode phase inverts this profile as shown in Figure~\ref{fig:figure4-decode-thrpt} showcasing lower throughput compared to prefill but shows a rising trend with increasing context. Figure~\ref{fig:figure4-gpu-hbm-util} shows high HBM bandwidth saturation ($\approx$85\% for 8B, $\approx$65\% for 405B in Figure~\ref{fig:figure6-llama}), while SM occupancy drops or becomes variable. This phase is \textit{Memory-Bound}. The arithmetic intensity collapses because the auto-regressive generation requires loading the entire model and KV cache to generate a single token. For reasoning workloads where $OSL \gg ISL$, the system spends the vast majority of wall-clock time in this inefficient, bandwidth-limited regime.

\begin{observationbox}
\textbf{Observation 7:} 
Reasoning workloads increase the fraction of time spent in decode, where arithmetic intensity is lower than prefill and performance is more constrained by KV movement and HBM bandwidth. As a result, high-FLOP compute units can remain underutilized even while latency remains high. This suggests that reasoning-serving optimization should prioritize KV locality, bandwidth efficiency, and decode scheduling rather than only maximizing peak compute utilization.
\end{observationbox}

\subsection{The Reasoning Cliff: KV Scaling}
\begin{figure*}
    \centering
    \begin{subfigure}[b]{0.6\textwidth}
        \centering
        \includegraphics[width=\linewidth]{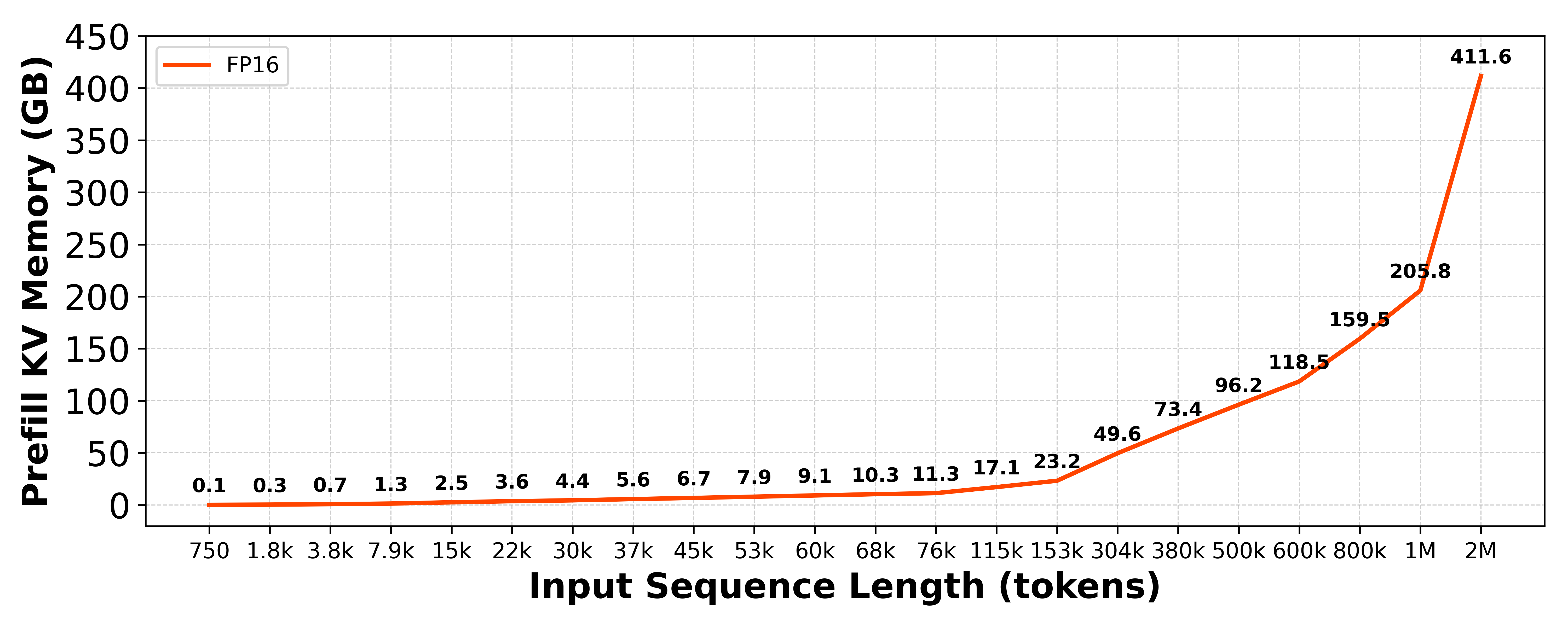}
        \caption{Prefill Phase KV Cache Utilization for varying ISL.}
        \label{fig:figure5-prefill}
    \end{subfigure}
    \hfill
    \begin{subfigure}[b]{0.6\textwidth}
        \centering
        \includegraphics[width=\linewidth]{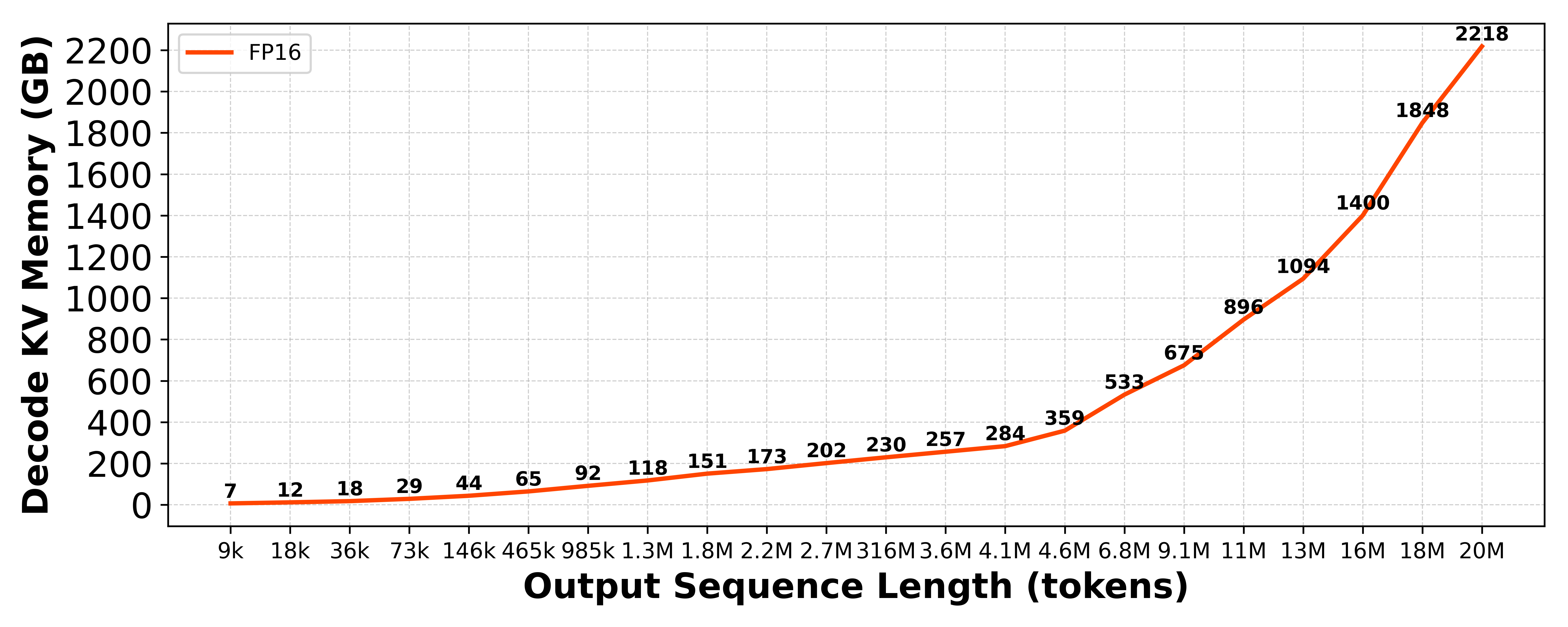}
        \caption{Decode Phase KV Cache Utilization for varying OSL.}
        \label{fig:figure5-decode}
    \end{subfigure}
    \caption{Analysis of Prefill and Decode Memory Requirements during AI Inference.}
    \label{fig:figure5-prefill-decode-kv-proj}
    \vspace{-10pt}
\end{figure*}

The severity of the memory bottleneck is dictated by the growth of the KV cache relative to the model weights. We characterize this scaling behavior in Figure~\ref{fig:figure5-prefill-decode-kv-proj} for the 8B model and Figure~\ref{fig:figure6-405B} for the 405B model.

\begin{observationbox}
\textbf{Observation 8:} 
The ``reasoning cliff'' occurs when the growing KV cache exceeds available HBM capacity and forces the scheduler to preempt, recompute, or reject additional work. For short-output workloads, this cliff typically appears later in decode. For reasoning workloads with long outputs, KV growth pulls the cliff earlier, sometimes limiting admission during prefill. Systems should therefore estimate future KV growth at admission time and reserve HBM capacity for long-running decode phases instead of admitting requests based only on current memory usage.
\end{observationbox}

\paragraph*{\bf Linearity vs. Capacity}
Figure~\ref{fig:figure5-decode} demonstrates that the decode-phase KV footprint grows linearly with OSL. For the 8B model, a 20M token output (simulating aggregate reasoning across a batch) demands over 2 TB of memory. While prefill KV also grows linearly (Figure~\ref{fig:figure5-prefill}), it is transient. The decode KV is persistent.
For the Llama-405B model (Figure~\ref{fig:figure6-llama-prefill-decode}), this scaling creates a "Reasoning Cliff." At a batch size of 1K with long context, the KV cache hits 100\% utilization during the decode phase. However, as the batch size increases to 4K or 5K, the saturation point shifts \textit{forward} in time as the cache exhausts during the prefill phase. This observation implies that for massive reasoning batches, the system cannot initialize requests before hitting the HBM wall.

\subsection{System Mitigation: Chunked Prefills and Preemption}
To survive the ``Reasoning Cliff'' mentioned above, the inference engine must dynamically degrade service, which introduces a negative dimension that is interesting to study. To this end, we analyze the vLLM scheduler's response to this pressure using the Llama-405B request analysis, as depicted in Figure~\ref{fig:figure6-llama-req-analysis}.

\begin{observationbox}
\textbf{Observation 9:} 
For reasoning workloads, the scheduler must regulate the number and length of concurrent decode streams to keep KV usage within the HBM envelope. This makes scheduling a memory-traffic shaping problem: HBM capacity bounds sustainable throughput, while HBM bandwidth bounds per-token latency. Practical schedulers should therefore incorporate KV-aware admission control, decode throttling, and priority policies that balance TTFT, TPOT, and preemption risk.
\end{observationbox}

\paragraph*{\bf Throttling Dynamics}
When the batch size reaches 5K (green line), the system cannot fit the full state in memory. The scheduler activates \textbf{Chunked Prefill}, processing only a subset of the prompt tokens per iteration.
\begin{itemize}[topsep=0pt,itemsep=0pt,leftmargin=12pt]
    \item \textbf{Queueing Behavior:} The ``Running'' curve (solid green) plateaus while the "Waiting" curve (dashed green) remains high. This indicates the system is serializing the admission of requests. Unlike the 500-batch case (blue), where requests enter and exit quickly, the 5K case forces the system into a "convoy" mode where new reasoning traces are only admitted as old ones finish and free their KV blocks.
    \item \textbf{Throughput-Latency Tradeoff:} By chunking prefills, the system prevents OOM crashes but introduces a "Start-Up Latency" for reasoning tasks. The GPU utilization remains high, but it is effectively stalling on memory capacity management rather than productive token generation.
\end{itemize}

\section{Discussion}
A forward-looking view of architecture design reveals a clear bifurcation between compute-intensive prefill and bandwidth-bound decode, motivating a decisive shift toward hardware disaggregation. Empirical characterizations underscore the fundamentally different bottlenecks of these phases, prompting next-generation systems to physically and logically decouple their execution paths to improve performance~\cite{d-matrix1}.

To meet the divergent requirements of large-scale inference, a tiered architecture that offloads the prefill phase to accelerators optimized for dense compute is optimal~\cite{dynamo}. High-TFLOP devices with HBM having moderate bandwidth can deliver superior arithmetic throughput while providing sufficient capacity and bandwidth for prompt ingestion, as long as the memory subsystem can sustain the higher core utilization. When capacity or bandwidth becomes a constraint, the system can bank on offloading to system memory and utilize multiple tiers together with high-speed interconnects such as NVLink. This arrangement balances prefill throughput against the cost and constraints of the GPU-HBM complex, achieving an efficient trade-off between compute, capacity, and interconnect provisioning.

In contrast, the decode phase is governed by memory bandwidth and capacity, making it a better fit for scalable, memory-centric hierarchies. Combinations of HBM with CXL~\cite{10.1145/3589013.3596678, cxl} and near-memory compute solutions, including 3D-stacked memory (e.g., SRAM-based designs from D-Matrix~\cite{d-matrix2}), alleviate the bandwidth pressure on KV reads. Operationally, decode benefits from explicit pooling and tiered offload across HBM → DDR/LP → CXL → NVMe, underpinned by fast links such as NVLink or optical to preserve fleet-wide latency and throughput. 3D-stacked options that add DRAM atop HBM at the same pin speed can raise effective bandwidth to KV accesses without widening I/O, reducing the need to proliferate external HBM stacks. The principal trade-off is that bandwidth scales faster than usable per-stack capacity due to thermal limits, yield, and stack height. If decode continues to demand both higher bandwidth and larger KV stores, deeper 3D stacks primarily address the bandwidth side while capacity grows more slowly and at higher cost and complexity. Balanced designs therefore pair 3D bandwidth tiers with pooled DRAM/CXL/NVMe for capacity, tied together with fast optical interconnects. Software can exploit this bifurcation by reserving high-bandwidth resources exclusively for token generation while routing prefill to cost-optimized tiers, using advanced runtime strategies to maximize efficiency.

High-bandwidth flash (HBF)~\cite{hbf} approaches that augment HBM with a NAND tier are a potential fit for both phases. For prefill, HBF can stream model weights effectively; for decode, it offers higher capacity for colder KV entries and weights. However, HBF introduces challenges: higher power, read/write asymmetry, lower bandwidth, and substantial software uplift. Prefill’s mixed read/write behavior makes it sensitive to this asymmetry, potentially depressing throughput even as NAND capacity helps accommodate growing input contexts. For decode, the extra capacity is beneficial for long output sequence lengths and reasoning-heavy workloads, but the higher bandwidth demands necessitate intelligent KV management and caching. In practice, funneling traffic through the limited HBM channels in HBF can create bottlenecks unless the software stack tightly orchestrates placement, prefetching, and eviction.

Agentic AI further amplifies memory-system pressure by transforming inference from a single long reasoning trajectory into a stateful, multi-step execution spanning both accelerator and CPU domains. A request may invoke planner, executor, verifier, retrieval, and tool-augmented model calls, each propagating intermediate context, reasoning traces, and active KV state across iterations. As a result, the memory footprint scales not only with model size and CoT length, but also with agent fan-out, branching depth, tool interaction frequency, and concurrent sessions.
Modern agentic architectures further rely on tightly coupled CPU–GPU execution, where CPUs sustain large numbers of independent agent environments, manage tool execution, and maintain per-agent state, queues, and retrieval data. Each agent incurs persistent DRAM overhead for context staging, tool outputs, sandbox execution, and vector data, placing host memory capacity and bandwidth directly on the execution path.
This shifts the bottleneck from isolated GPU memory capacity to a system-wide problem: HBM must sustain high KV residency and reuse, while host DRAM actively participates in context management, data staging, and CPU-side processing. Compared with single-shot inference, agentic workloads create a multiplicative demand across HBM and DRAM, stressing both per-GPU limits and rack-level scheduling, motivating KV-aware placement, migration, and tiered memory management across HBM, host, CXL, and storage tiers under strict TTFT, throughput, and ms/token SLAs.

Given sublinear performance scaling with model size and tight KV capacity constraints, future systems should move beyond monolithic, GPU-local KV caches and adopt explicit control over KV placement and migration. Offloading, pooling, and hierarchical tiering across HBM and host memory enable proactive eviction, compression, and reuse of pre-generated blocks, materially reducing time to first token. This memory flexibility allows heterogeneous deployments that combine small, distilled models for high-throughput speculative decoding with large foundation models for complex queries. To bind these disaggregated components, next-generation interconnects especially NVLink and optical links are essential across both prefill and decode paths, delivering the low-latency, high-throughput data movement required across tiers and between accelerators.

\section{Conclusions}
This work presented a systematic characterization of GPU-based inference for reasoning-centric LLMs, evaluating models from 8B to 671B parameters across diverse parallelization strategies. Our results highlight a fundamental divergence in scaling behavior: Data Parallelism achieves near-linear throughput for small models but faces hard capacity limits from KV-cache saturation, while Tensor Parallelism is essential for ultra-large models yet remains bottlenecked by memory bandwidth and synchronization costs. A key contribution of this study is the quantification of the "performance cliff" in reasoning workloads. We demonstrated that the extended context lengths required for multi-step reasoning exacerbate the disparity between the compute-bound prefill phase and the memory-bandwidth-bound decode phase. As sequence lengths grow, the KV cache aggressively consumes HBM capacity, leading to rapid throughput degradation and request throttling once memory limits are exceeded. These insights suggest that standard monolithic scaling is reaching its limits for reasoning-dense applications. Future inference systems must prioritize architectural disaggregation, effectively decoupling the prefill and decode phases to optimize for their respective compute and bandwidth requirements. Furthermore, realizing the full potential of frontier-scale models will require innovations in tiered memory management and hardware-software co-design to mitigate the severe capacity and bandwidth constraints identified in this analysis.

\section*{Acknowledgements}
This work was supported in part by the U.S.\ Department of Energy under Contract DE-AC02-06CH11357, as well as the National Science Foundation (NSF), Office of Advanced Cyberinfrastructure, under grants CSSI-2411386 and CSSI-2514056.

\bibliographystyle{IEEEtranS}
\bibliography{refs}

\end{document}